\documentclass[12pt]{article}

\usepackage{caption,hyperref,url}
\usepackage[pdftex]{graphicx}
\usepackage{amsmath,amssymb,bm}
\usepackage{siunitx}
\usepackage{xcolor}
\usepackage{lineno}
\usepackage{subfiles}
\usepackage{titling}  
\usepackage{changepage}
\usepackage{multirow}
\usepackage{enumerate}
\usepackage[shortlabels]{enumitem}

\graphicspath{{./figures/}}

\title{Probing dynamics of a two-dimensional dipolar spin ensemble using single qubit sensor}

\author
{Kristine Rezai$^{1,3}$, Soonwon Choi$^{2}$, Mikhail D. Lukin$^{1}$, Alexander~O.~Sushkov$^{3,4,5}$
\\
\\
\normalsize{$^{1}$Department of Physics, Harvard University, Cambridge, MA 02138, USA}\\
\normalsize{$^{2}$Department of Physics, Massachusetts Institute of Technology, Cambridge, MA 02142, USA}\\
\normalsize{$^{3}$Department of Physics, Boston University, Boston, MA 02215, USA}\\
\normalsize{$^{4}$Department of Electrical and Computer Engineering, Boston University,}\\
\normalsize{Boston, MA 02215, USA}\\
\normalsize{$^{5}$Photonics Center, Boston University, Boston, MA 02215, USA}\\
\normalsize{$^*$To whom correspondence should be addressed; e-mail: asu@bu.edu}\\
}

\date{}

\begin{document}


\maketitle
\begin{abstract}
Understanding the thermalization dynamics of quantum many-body systems at the microscopic level is among the central challenges of modern statistical physics.
Here we experimentally investigate individual spin dynamics in a two-dimensional ensemble of electron spins on the surface of a diamond crystal.
We use a near-surface NV center as a nanoscale magnetic sensor to probe correlation dynamics of individual spins in a dipolar interacting surface spin ensemble.
We observe that the relaxation rate for each spin is significantly slower than the na\"ive expectation based on independently estimated dipolar interaction strengths with nearest neighbors and is strongly correlated with the timescale of the local magnetic field fluctuation.
We show that this anomalously slow relaxation rate is due to the presence of strong dynamical disorder and present a quantitative explanation based on dynamic resonance counting.
Finally, we use resonant spin-lock driving to control the effective strength of the local magnetic fields and reveal the role of the dynamical disorder in different regimes.
Our work paves the way towards microscopic study and control of quantum thermalization in strongly interacting disordered spin ensembles.
\end{abstract}

Quantum thermalization connects statistical physics with unitary quantum mechanics.
Recent technological developments in quantum information science have enabled detailed studies of isolated quantum systems, revealing a variety of novel phenomena such as the role of entanglement in thermalization~\cite{Kaufman2016,Lukin2019,Rispoli2019}, the localization in the absence of strong disorder~\cite{Morong2021}, non-equilibrium phases~\cite{Zhang2017,Choi2017,Keesling2019,Rispoli2019,Ebadi2021,Kyprianidis2021}, and quantum many-body scarring~\cite{Turner,Bluvstein2021,Kao2021}.
Despite this progress, one of the key open questions in this field is the nature of microscopic relaxation dynamics in presence of dynamical disorder and long-range interactions~\cite{Cardellino2014}.
The majority of existing studies focus on ensemble measurements in systems dominated by static disorder~\cite{Alvarez2015,Wei2018,Smith2016,Choi2016,Kucsko2018,Schreiber2015}. However ensemble averaging can conceal important features of microscopic dynamical evolution~\cite{Feldman1996,Dobrovitski2008}, and in many noisy real-world quantum systems disorder is dynamic, exhibiting non-trivial time-dependence.
Another important factor is system dimensionality and its interplay with the distance scaling of interactions. The long-range $1/r^3$ dipolar spin interaction implies that a two-dimensional system should be localized according to the single-particle Anderson model~\cite{Anderson1958}, but delocalized according to the many-body interacting treatment~\cite{Burin2006,Yao2014,Gong2017a,Orioli2018,Gopalakrishnan2016}. In addition to fundamental interest, understanding the dynamics of two-dimensional systems is important for quantum sensing applications, since such systems can be positioned in close proximity to a sensing target~\cite{Sushkov2014,Davis2021}.

We experimentally investigate spin transport dynamics of a two-dimensional ensemble of randomly-positioned spin-1/2 qubits with long-range magnetic dipolar interactions~\cite{Davis2021}.
Na\"ively, one could expect that the spin exchange (flip-flop) component of dipolar interactions limits the lifetime of local polarization of spins.
Our observation, however, reveals a surprising finding that the spin lifetime is, in fact, significantly longer than independently-estimated interaction strengths.
We attribute this anomalously-slow spin relaxation dynamics to the interplay between interactions and strong time-dependent local disorder, created by hyperfine fields of proximal nuclear spins.
We control the effective strength of this disorder and observe the corresponding scaling of the relaxation rate.
We present a theoretical model, based on dynamic resonance-counting arguments, which is in quantitative agreement with our experimental observations and reveals a universal scaling collapse of our data for different values of disorder strengths and correlation times.

Our experimental platform is the ensemble of paramagnetic two-level systems on the surface of a diamond crystal, fig.~\ref{fig:1} (a)~\cite{Grotz2011,Grinolds2014,Tetienne2018}.
These surface spins are associated with 
electron spin $S=1/2$ impurities that are not optically active. Their exact nature is subject to ongoing investigation, but they are likely localized defect surface states
~\cite{Sangtawesin2019,Stacey2019}.
Due to faster decoherence of shallow NV centers~\cite{Rosskopf2014,Romach2015,Myers2014,Myers2017,Bluvstein2019}, these surface spins have been considered to be deleterious and significant effort has gone into treating and engineering diamond surfaces to minimize their density~\cite{FavarodeOliveira2017}. 
However, with proper quantum control, they
can be turned into a useful resource.
The surface spins can be coherently manipulated and measured, and can be used as the so-called quantum reporters that probe and report the local magnetic environment. For example, by addressing a single surface spin, it is possible to detect and localize proximal single proton nuclear spins on the diamond surface under ambient conditions~\cite{Sushkov2014}.

The dynamics of individual surface spins are measured by a single near-surface NV center, acting as a nanoscale sensor of magnetic fields created by the surface spins. NV centers are addressed using a confocal microscopy setup, combined with radiofrequency (RF) spin drive fields that are delivered via a transmission line fabricated on a glass coverslip. 
The surface spin transitions can be addressed with RF pulses delivered in the same manner. 
The 2.87 GHz zero-field splitting of the NV center enables independent addressing of the NV spin and the surface spin transitions by using different resonant RF tones at a given bias magnetic field, aligned with the NV axis (z-axis), fig.~\ref{fig:1}(a), inset.
The magnetic dipole coupling between the NV center and the surface spin ensemble is characterized using a double electron-electron resonance (DEER) sequence, fig.~\ref{fig:1}(b).
The NV center spin-echo decays on the timescale $T^{(nv)}_2$, but when a $\pi$-pulse flips the surface spins simultaneously with the NV spin, the NV spin echo collapses on a timescale that depends on the strength of the dipolar field created by the surface spins near the NV center. 
Because the magnetic dipole interaction is long-range, the NV center is, in general, coupled to multiple surface spins, with coupling strengths dependent on locations of surface spins on the diamond surface. 
The oscillations present in the data at short time (fig.~\ref{fig:1}(b) red points) indicate that there is one ``central'' surface spin, whose dipolar coupling strength to the NV center $k_{max}$ dominates over other surface spins.
Recent experimental studies have shown that after certain surface treatments, some of the surface spins may be mobile, likely changing their positions under green laser light illumination~\cite{Dwyer2021}. We have verified that in our experiments the central surface spin remains in place, even when subjected to green laser light illumination of up to 50 $\mu$s~\cite{som}.
In fact, the central spin position is stable for the experiments carried over the timescale of months.
This is likely due to the combination of small photo-ionization cross-section and steric protection by chemical surface groups~\cite{Dwyer2021}.

Our experiments take place at room temperature, with the bias magnetic field on the order of 1000~G. Therefore the surface spin ensemble is effectively at infinite spin temperature. Nevertheless, we study the dynamics of an individual central surface spin $S_j$, by measuring its spin autocorrelation functions.
Quantum logic gates between the NV center and the central surface spin correlate the NV and the central spin evolution, via their magnetic-dipole interaction, fig.~\ref{fig:2}(a)~\cite{Laraoui2013}.
Each measurement consists of a sequence of RF pulses, applied to the NV and surface spins, as well as NV optical spin polarization and readout steps. The RF pulse sequences include two probe segments (CNOT gates), in which the NV center probes the quantum state of the surface spin ensemble, separated by a time interval, in which this state can evolve with control pulses applied to the surface spins (fig.~\ref{fig:2}(a), grey box).
The interpulse spacing $t_{NV}$ of the probe segments is tuned to the timescale of the dipolar interaction between the NV center and the central surface spin so that the final NV center state is contingent on whether the z-projection of the central surface spin changes between the gates.
Applying this method, the NV center acts as a probe of central surface spin autocorrelation functions $\langle S^{x,y,z}_j(t)S^{x,y,z}_j(0)\rangle$, where the time $t$ and the measured spin projection depend on the pulses applied during the evolution interval. We performed measurements on 11 separate NV center-surface spin systems~\cite{som}.  

The Ramsey measurement probes fluctuations of the local magnetic field at the central surface spin site (fig.~\ref{fig:2}(b) blue). The Hahn echo sequence decouples the central surface spin from low-frequency fluctuations, which extends the coherence time to $T_2$ (fig.~\ref{fig:2}(b) red). Further decoupling can be achieved with higher-order dynamical decoupling experiments, such as XY-4 (fig.~\ref{fig:2}(c) blue). However we observe that this does not further increase the coherence time, indicating the presence of a separate decoherence mechanism. The presence of dipolar interactions with other surface spins motivates the MREV-8 experiment, which decouples dipolar interactions. Indeed, we observe that the coherence time in an MREV-8 experiment is extended, compared to the Hahn echo and XY-4 (fig.~\ref{fig:2}(c) red curve), showing that the dynamics of surface spins are strongly affected by dipolar interactions between them.

In order to understand our observations quantitatively, we consider the effective Hamiltonian in the rotating frame of a single surface spin $S_j$:
\begin{equation}
    H_j = \hbar \gamma_e B^z_j(t) S_j^z + \sum_i \frac{\hbar^2 \gamma_e^2}{r_{ij}^3}(1-3\cos^2\theta_{ij})[S_i^z S_j^z - \frac{1}{4}(S_i^+ S_j^- + S_i^- S_j^+)],
    \label{eq:1}
\end{equation}
where $\hbar$ is the reduced Planck constant, $\gamma_e = 2\pi\times2.80$ MHz/G is the electron gyromagnetic ratio, $B^z_j(t)$ is the fluctuating magnetic field at the site of the surface spin, $r_{ij}$ is the distance between the central surface spin $j$ and a different surface spin $i$, and $\theta_{ij}$ is the angle between the direction of the applied external magnetic field and the vector $\bm{r}_{ij}$.
In eq.~\eqref{eq:1} we do not include the dipolar interaction between the NV center and the central surface spin. By running experiments that vary the NV center spin state during the surface spin evolution period, we find that this interaction does not affect central surface spin evolution within our experimental uncertainty~\cite{som}.

The strengths of the dipolar interaction terms in eq.~\eqref{eq:1} vary over different pairs of spins, due to their random positions on the surface. However the $1/r^3$ distance dependence and the 2D nature of the surface spin ensemble imply that it is the proximal surface spins that dominate the spin echo decoherence. We use the spin echo data to extract the strength of the central spin interaction with these proximal spins: $J_1 = 1/T_2 = (0.71 \pm 0.05)\,\mu$s$^{-1}$. 

Let us consider the origin of the fluctuating local magnetic field $B^z_j$.
All measurements were preformed with the diamond surface submerged in  deuterated glycerol. This reduced the density of proton nuclear spins near the surface. However, as observed in several other studies of near-surface NV centers, an intrinsic $\sim$ 1 nm thick layer of surface water and hydrocarbons contains a high density of proton nuclear spins~\cite{Staudacher2015,Loretz2014,DeVience2015}. 
The periodic features that appear at odd multiples of proton Larmor period in the spin echo, XY-4, and MREV-8 data indicate that these protons are the dominant source of the local field $B^z_j$. We model the dynamics of $B^z_j$ by parametrizing its power spectrum as a combination of two terms:
\begin{equation}
    V(\omega) = \int_{-\infty}^{\infty}\langle B^z(t) B^z(0) \rangle e^{-i\omega t}\,dt=2\frac{W^2}{\gamma_e^2}\left(\frac{\tau}{\omega^2\tau^2+1}+\frac{5}{9}\frac{\tau}{(\omega-\omega_L)^2\tau^2+1}\right).
    \label{eq:2}
\end{equation}
The first term, centered at zero frequency, quantifies slow fluctuations with strength $W$ and correlation time $\tau$, due to proton spin projections along the bias magnetic field~\cite{som}. The second term quantifies fluctuations near the proton Larmor frequency, due to transverse proton spin projections (fig.~\ref{fig:2}(d)).
We simultaneously fit the Ramsey, spin echo, XY-4, and MREV-8 experimental data shown in fig.~\ref{fig:2}, with $J_1$, $W$, and $\tau$ as fit parameters~\cite{som}. For this NV-surface spin system, we extract $W$ = (4.40 $\pm$ 0.38) $\mu$s$^{-1}$ and $\tau$ = (14.6 $\pm$ 4.2) $\mu$s. 
We note that the decoherence in the Ramsey sequence is significantly faster than that of the spin echo, indicating that $W$ is greater than $J_1$. The decoherence of the MREV-8 data is dominated by the low-frequency noise component in eq.~\eqref{eq:2}.

Having characterized the local environment of the central surface spin, we study the spin flip-flop transport dynamics in this system by measuring the autocorrelation of the spin component along the applied magnetic field: $\langle S_j^z(t)S_j^z(0)\rangle$, fig.~\ref{fig:3}(a). The presence of the $S^+S^-$ flip-flop terms in the dipolar interaction in eq.~\eqref{eq:1} motivates an expectation that this single-spin autocorrelation would decay over the timescale $T_z$, which is on the order of $1/J_1=T_2$. In contrast, we consistently observe remarkably slow decay of the longitudinal correlation, which persists over a timescale $T_z$, in this instance $\sim 30\times$ slower than the dipolar interaction timescale $T_2$, fig.~\ref{fig:2}(e). We emphasize that these observations are especially unexpected for individual spin measurements, in contrast with experiments probing a macroscopic spin ensemble (such as a bulk magnetic resonance measurement), where the total spin z-projection is unaffected by flip-flops within the ensemble.

We attribute such anomalously slow relaxation dynamics to the presence of strong disorder. 
To quantify its effect, we consider a theory model based on an effective single-particle resonance counting.
In this model, each spin $S_j$ experiences a time-dependent energy shift owing to the combination of two effective sources of disorder: i) the extrinsic on-site disordered magnetic field $B_j^z(t)$ of typical strength $\sim W$ and ii) the intrinsic local field of strength $\sim J_1$ arising from the Ising component of the dipolar interactions among surface spins. These two components have distinct correlation times and strengths. 
To quantify the combined dynamical disorder, we introduce the effective disorder strength $W_e = \sqrt{W^2+J_1^2}$ and the effective disorder correlation time $\tau_e$, defined via $\sqrt{1/\tau_e} = (W\sqrt{1/\tau}+J_1\sqrt{1/T_z})/W_e$~\cite{som}.
We assume spins exchange their polarization via dipolar interactions when a pair of spins $i,j$ becomes ``resonant'', i.e. when the difference between their spin energy shift is smaller than the flip-flop interaction between them, fig.~\ref{fig:3}(d). 
Otherwise, energy conservation blocks flip-flops~\cite{Kucsko2018}.
The spin relaxation dynamics is probed by estimating the probability for a central spin to flip-flop with its neighbor as a function of time.
This model predicts that for a 2D dipolar spin system, after averaging over random positioning of spins, the central spin autocorrelation $\langle S_j^z(t)S_j^z(0)\rangle$ decays with an approximate functional form $\exp{(-(t/T_z)^{2/3})}$~\cite{som}, where $T_z = \kappa \tau_eW_e/J$. The numerical constant $\kappa$ is of order unity, and $J$ is the average dipolar interaction strength over the spin ensemble.
We compare this prediction with our experimental data by scaling the time axis of the autocorrelation data by independently estimated $\tau_eW_e$, and observe that the data sets for the 7 different NV-central spin systems collapse towards a universal curve, fig.~\ref{fig:3}(b). 
Plotting the individual relaxation times $T_z$ versus the product $\tau_eW_e$ reveals the linear relationship predicted by our resonance counting, with the best-fit value $\kappa$ = (0.31 $\pm$ 0.14) and J = (0.57 $\pm$ 0.25) $\mu$s$^{-1}$, consistent with the interaction strengths extracted from the spin echo and XY-4 data sets, fig.~\ref{fig:3}(c).

To further investigate the role of disorder, we control the magnitude of the effective disorder by spin-lock driving, on resonance with the surface spin Larmor frequency. The resonant drive along the y-axis of the rotating frame adds to the Hamilonian \eqref{eq:1} the additional term $H_d = \hbar \Omega S^y$, where $\Omega$ is the drive Rabi frequency.
We measure the dynamics of the central surface spin projection $S_j^y$ along the driving field, fig.~\ref{fig:4}(a). This is equivalent to $T_{1\rho}$ measurement in NMR~\cite{Abragam1961,Slichter1978}.
The relaxation at small $\Omega$ is dominated by the local magnetic field noise $B_j^z(t)$, which can flip $S_j^y$. In this regime, the relaxation rate calculated using the noise model in eq.~\eqref{eq:2} is consistent with our measurements (fig.~\ref{fig:4}(b), black line).
At large drive strength $\Omega$, spin flips due to $B_j^z(t)$ are suppressed, such that the strength of the effective extrinsic disorder scales as $W^2/(\sqrt{2}\Omega)$. 
Consequently, the disorder experienced by individual spins is dominated by the Ising component of dipolar interactions along the dressed quantization axis $S_j^y$. The strength of the latter is reduced to the half of the original Ising interaction along $S_j^z$ in the undriven case~\cite{Kucsko2018}.
Thus, systematic comparisons of $T_{1\rho}/2$ and $T_z$ enable us to understand the interplay between the extrinsic and intrinsic disorder fields.
Shown in fig.~\ref{fig:4}(c), we find that the relaxation time becomes relatively faster by suppressing the extrinsic disorder when $W$ is sufficiently large.
However, for samples with small or intermediate $W$, the difference in $T_{1\rho}$ and $T_z$ is not substantial, suggesting that the Ising interaction constitutes the dominant source of disorder in this regime.
Interestingly, even in this regime, the relaxation rate is approximately a factor of 20 slower than the dipolar interaction scale $J_1$. 
This implies that, even in the absence of extrinsic on-site disorder, the intrinsic disorder associated with random positioning of spins is sufficient to strongly suppress spin transport in 2D spin ensemble~\cite{som}. 

A full many-body theoretical treatment of a random dipolar-interacting two-dimensional spin ensemble may be needed to quantitatively predict the relaxation timescale in the absence of extrinsic disorder.
A mean-field approach may also provide an approximate solution~\cite{Graser2021}.
It is natural to inquire if a many-body localized phase could be observed for this system.
While we do not observe localization directly 
within the accessible parameter range, we note that theoretical analyses of localization physics typically consider quasi-static disorder models, where the value of on-site field strength does not change on the timescale of a single measurement.
Quasi-static disorder generally slows down relaxation as shown in our experiments, but our model demonstrates that time-dependent disorder may speed it up, for a subset of values of $W$ and $\tau$, by allowing previously off-resonant spins to become resonant.
Achieving finer control over disorder parameters may allow a detailed single-spin-resolution study of various aspects of localization physics in the present system.  

Our observations open the door for in-depth explorations of quantum  dynamics in two-dimensional spin systems. In particular, they indicate that systems of interacting spin qubits with long coherence times can be created and manipulated under ambient conditions at room temperature.  Our approach can be used used for the controlled generation of entanglement among spins on the diamond surface with potential applications to quantum sensing and magnetic resonance imaging of single molecules.

\textbf{Acknowledgements:}

The authors acknowledge discussions with Timo Gr\"{a}\ss{}er, G\"{o}tz Uhrig, Elana Urbach, Tamara Sumarac, Emma Rosenfeld, Bo Dwyer, Harry Zhou, and Oleg P. Sushkov.
This work was supported by the National Science Foundation grant PHY-2014094, NSF, CUA, ARO MURI, and Moore Foundation.

\bibliography{library,library_a} 

\newpage

\begin{figure}[h]
\begin{center}
\includegraphics[width=0.9\textwidth]{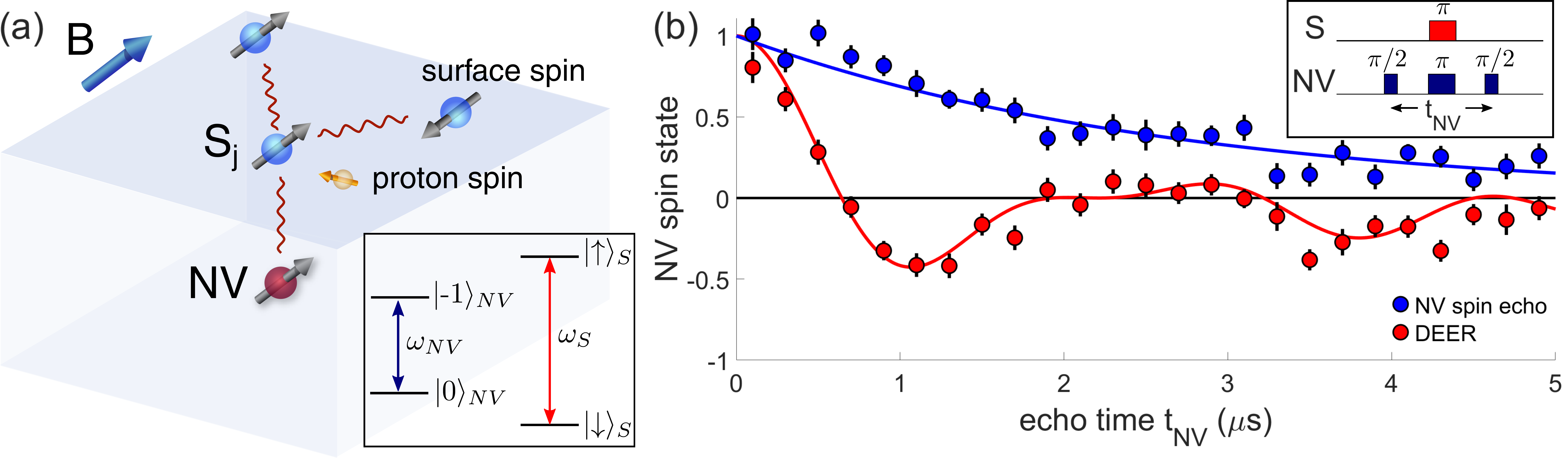}
\end{center}
\caption{\linespread{1.0}\fontsize{9}{12}\selectfont
Characterization of a representative NV center - surface spin system.
(a) The experimental platform consists of a single, shallow NV center strongly coupled to one surface electron spin $S_j$.  Other electron spins and proton nuclear spins reside on the surface of the diamond.  An external magnet splits the electron and nuclear spin states and allows independent manipulation of NV center and surface spins at their corresponding magnetic resonance frequencies.
Inset: NV center and surface spin level structure.  NV center transition is addressed at angular frequency $\omega_{NV}$ while surface spin transition is addressed at $\omega_S$.
(b) NV center Hahn echo (blue) and double electron-electron resonance (DEER) (red) measurements.  The oscillations present in the DEER data indicate that the magnetic dipole interactions of this NV center with the surface spin ensemble are dominated by a central surface spin with the coupling strength $k_{max}$ = (4.90 $\pm$ 0.57) $\mu$s$^{-1}$ (red line fit). Inset: the DEER pulse sequence.
}
\label{fig:1}
\end{figure}

\begin{figure}[h]
\begin{center}
\includegraphics[width=1\textwidth]{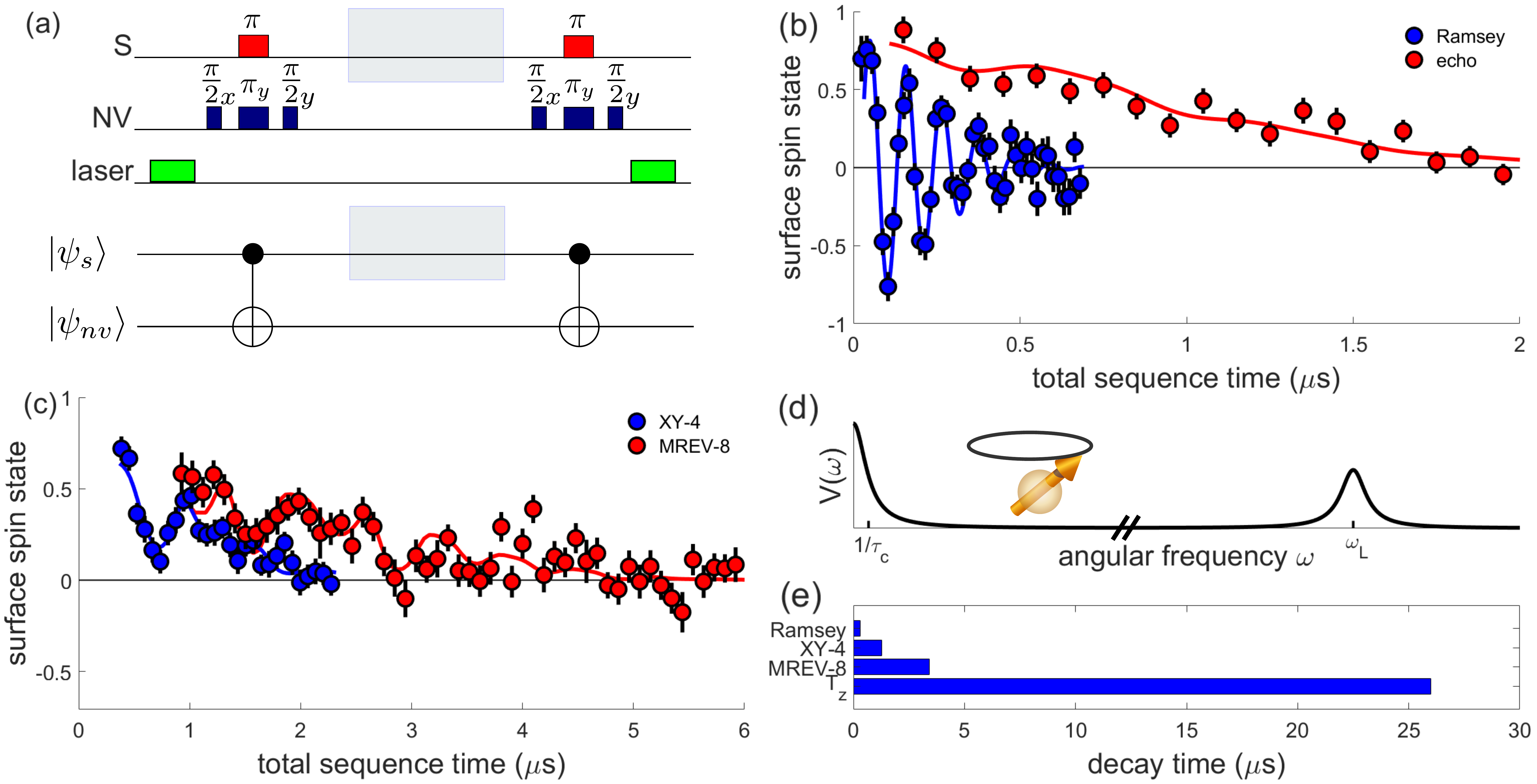}
\end{center}
\caption{\linespread{1.0}\fontsize{9}{12}\selectfont
Characterization of the local magnetic environment of a central surface spin.  
(a) Top: the correlation spectroscopy pulse sequence. In our experiments the NV center interpulse spacing is set such that the NV center is primarily sensitive to the central surface spin autocorrelation. Manipulating the surface spins during the gray shaded region allows us to measure different central surface spin correlation functions. The complete pulse sequences for all measurements are described in SI.
Bottom: Quantum logic gate representation of the correlation spectroscopy pulse sequence. The two CNOT gates are applied, such that the final state of the NV center is contingent on whether the z-projection of the central surface spin changed during the time between the gates.
(b) Surface spin detuned Ramsey (blue) and Hahn echo (red) measurements of $\langle S^{x}_j(t)S^{x}_j(0)\rangle$ transverse central spin autocorrelation.
The Hahn echo decay time is $T_{2}$ = (1.41 $\pm$ 0.11) $\mu$s and the Ramsey decay time is $T_2^*$ = (0.32 $\pm$ 0.02) $\mu$s. Solid lines are fits using the model in eqs.~(\ref{eq:1},\ref{eq:2}). The oscillations in the Ramsey data are due to the 9.2 MHz detuning from the surface spin Larmor frequency.
(c) Surface spin XY-4 measurement (blue) and MREV-8 measurement (red). The MREV-8 measurement is performed within a spin echo (see SI for full pulse sequences). Solid lines are fits using our model. The oscillations present in both measurements are due to proton nuclear spin precession.
(d) Nuclear spin bath noise power spectrum V($\omega$), with peaks at zero frequency and proton Larmor frequency $\omega_L$.
(e) The full characterization of relaxation timescales for one of the central surface spin systems, including the Ramsey, XY-4, MREV-8, and $S^z$ autocorrelation measurements.
}
\label{fig:2}
\end{figure}

\begin{figure}[h]
\begin{center}
\includegraphics[width=0.6\textwidth]{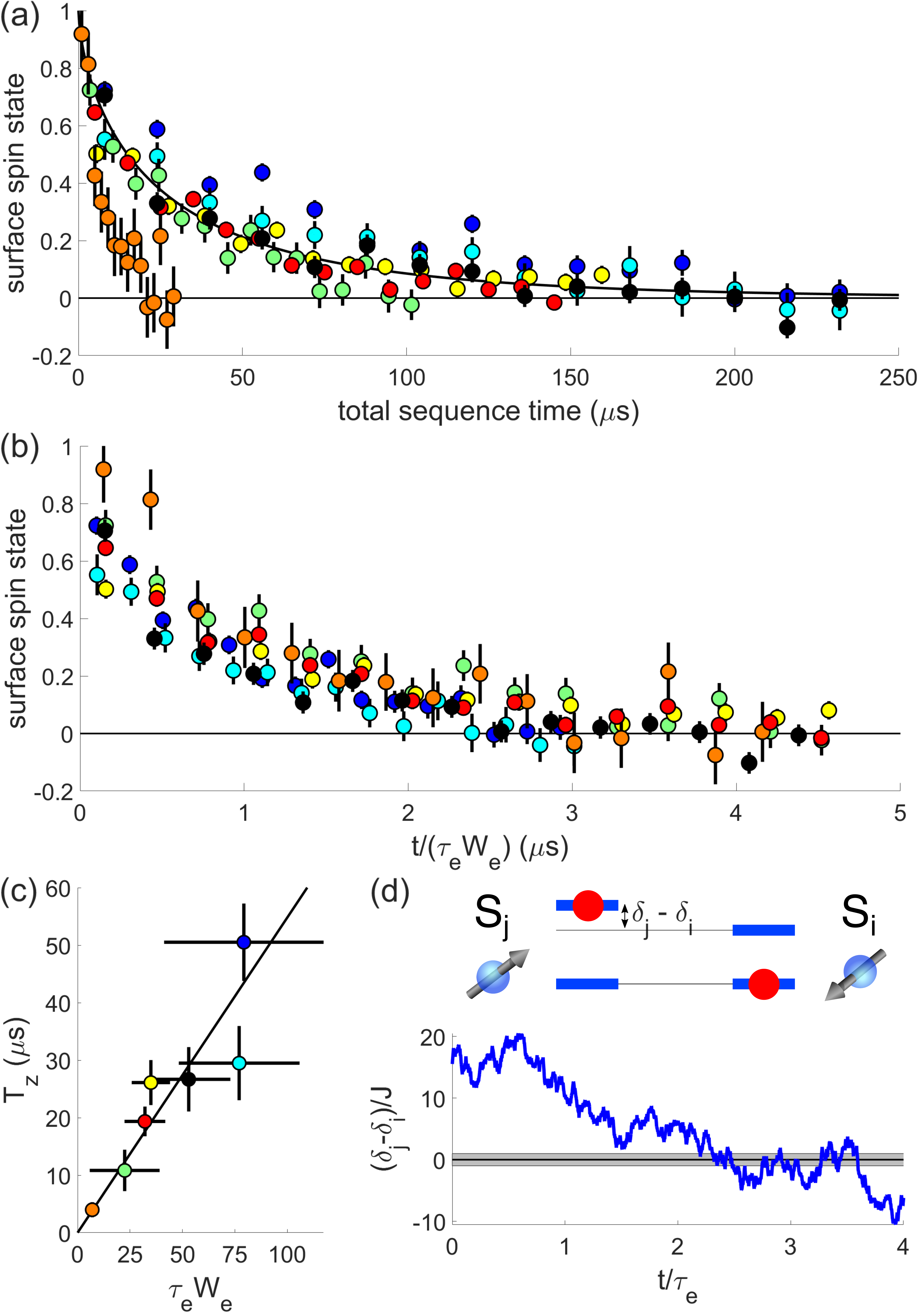}
\end{center}
\caption{\linespread{1.0}\fontsize{9}{12}\selectfont
Measurements of central surface spin $\langle S^{z}_j(t)S^{z}_j(0)\rangle$ autocorrelation.
(a) $S^z$ autocorrelation measurements for 7 different central surface spin systems. The black line is a stretched-exponential fit to one of the data sets, from which we extract the relaxation time $T_z$.
(b) $S_z$ autocorrelation measurements with the delay time re-scaled by the product $\tau_e W_e$ of the effective disorder correlation time and disorder width, independently measured for each surface spin system. We observe a collapse of the 7 data sets to a universal stretched-exponential decay curve, in agreement with our dynamical disorder model.
(c) S$_z$ autocorrelation decay time plotted as a function of the product $\tau_e W_e$, with the linear fit shown by the black line.
(d) A schematic of the spin hopping process in our model, showing a central surface spin $S_j$ and another surface spin $S_i$. Each spin state is shifted by time-dependent on-site detuning $\delta$ (blue line), due to the local dynamical disorder created by nuclear spin bath noise and Ising dipolar interaction terms. 
The spins can flip-flop if their detunings differ by less than the strength $J$ of the dipolar interaction between them (grey band).
}
\label{fig:3}
\end{figure}

\begin{figure}[h]
\begin{center}
\includegraphics[width=.7\textwidth]{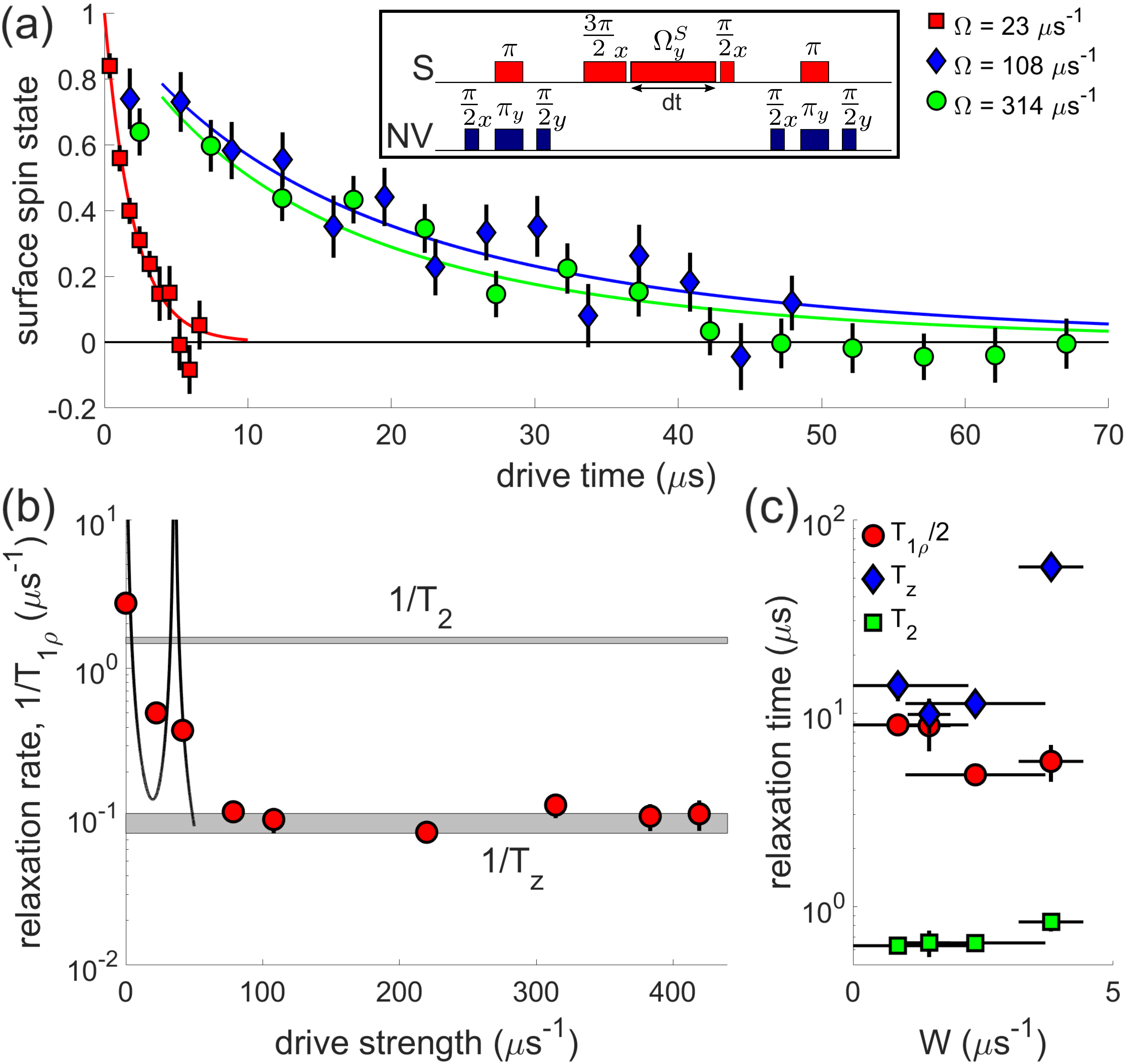}
\end{center}
\newpage
\caption{\linespread{1.0}\fontsize{9}{12}\selectfont
Evolution of central surface spin under spin-lock driving. 
(a) Measurements of $\langle S^{y}_j(t)S^{y}_j(0)\rangle$ autocorrelation under spin-lock driving for different drive strengths. Stronger driving results in better decoupling from the nuclear spin bath magnetic noise and longer $T_{1\rho}$ relaxation time. For drive strength $\Omega\gtrsim 50\,\mu$s$^{-1}$, relaxation rate is independent of $\Omega$. Lines show stretched exponential decay fits. Inset: radio frequency NV and surface spin pulses; green laser polarization and readout pulses applied before and after this sequence of RF pulses.
(b) Spin-lock driving relaxation rate as a function of drive strength for the surface spin system presented in (a), along with the corresponding decay rates for Hahn echo (1/$T_2$) and $S^z$ autocorrelation (1/$T_z$) experiments, shown by the gray bands. The point at $\Omega = 0$ corresponds to the decay of the Ramsey coherence (1/$T_2$*). The measurements performed with drive strength below 50 $\mu$s$^{-1}$ are consistent with direct relaxation due to the nuclear spin bath noise with spectrum V($\omega$). Relaxation rates for drive strengths $\Omega\gtrsim 50\,\mu$s$^{-1}$ are independent of $\Omega$ and consistently much slower than the dipolar interaction scale 1/$T_2$.
(c) Comparison between different autocorrelation decay timescales for 4 different surface spin systems. Red circles, labeled $T_{1\rho}/2$, show the decay time under spin-lock at strong driving.  As the dynamical on-site disorder strength W increases, the $T_z$ autocorrelation decay time increases compared with the autocorrelation decay at strong driving with suppressed disorder ($T_{1\rho}/2$).  Both of these decays are much slower (by a factor of $\approx 30$) than the dipolar interaction scale, given by $T_2$.
}
\label{fig:4}
\end{figure}

\clearpage
 


\renewcommand{\thefigure}{S\arabic{figure}}
\renewcommand{\thetable}{S\arabic{table}}

\title{Supplemental Material for \\ ``Probing dynamics of a two-dimensional dipolar spin ensemble using single qubit sensor''}


\maketitle


\appendix
\clearpage
\tableofcontents


\begin{adjustwidth}{-50pt}{-50pt}
	
	\section{Materials and Setup}
	
	\subsection{Diamond Sample}
	
	Nitrogen implantation was done on a $^{12}$C enriched $\{$001$\}$ diamond surface at an implantation energy of 3 keV.  After annealing, the diamond was cleaned in a 3-acid mixture (equal volumes of concentrated H$_2$SO$_4$, HNO$_3$, and HClO$_4$) for 3 hours.  All measurements in the main text were performed with the diamond placed in deuterated glycerol (CDN Isotopes D-0302) and under ambient conditions.
	
	
	\subsection{Optical Setup}
	
	A home built scanning confocal microscope optically addresses and reads out single NV centers.  The diamond was placed NV side down on a RF transmission line, and an inverted Nikon CFI Apo TIRF objective (NA $=$ 1.49) was positioned under the stripline.  A piezoelectric scanner (Physik Instrumente P-721 PIFOC) was used to control the height of the objective, and a closed-loop scanning galvanometer (Thorlabs GVS012) was used to direct the lateral position of the laser beam focus.  Optical excitation was provided by a 532 nm laser (Opto Engine MLL-FN-532-200mW) gated by an acousto-optic modulator (Isomet 1250C) in a double pass configuration.  NV center fluorescence passes through a 842 nm short pass filter (Semrock FF01-842/SP-25), a 594 nm long pass filter (Semrock BLP01-594R-25), and a 532 nm notch filter (Semrock NF03-532E-25) before collection by a single-photon counter (Excelitas Technologies SPCM-AQRH-14-FC).  The single-photon counter and acousto-optic modulator were gated using TTL pulses produced by a 500 MHz PulseBlasterESR-PRO pulse generator from Spincore Technologies.  A typical photon counting rate of 20 kHz was observed, and a photon counter acquisition window of 900 ns was used.

	\subsection{Radiofrequency Setup}
	
	Two Stanford Research SG384 signal generators were used to deliver RF tones to address the NV center and surface spin transitions.  
	IQ modulation was used to control the x and y quadrature amplitude of the signal generator driving the NV center.  
	The output of the signal generator driving the surface spins is fed into a 90 degree phase shifter (RF-Lambda RFHB01G04GVT).  
	The two outputs of the 90 degree phase shifter were each fed into 180 degree phase shifters (Pulsar Microwave JSO-10-47/3S).  
	Microwave switches (MiniCircuits ZASWA-2-50DR+), controlled by TTL pulses produced by the pulse generator, gated the RF drives.  
	The zero, 90 degree, 180 degree, and 270 degree phase shifted RF pulses that drive the surface spins were combined using a SigaTek SP51400 combiner.  
	The RF signals were power boosted by MiniCircuits amplifiers ZHL-5W-422+, ZHL-30W-252-S+, and ZHL-16W-43-S+, combined using MCLI PS2-134, and fed into a 50$\Omega$-terminated RF transmission line with the diamond sample on top. 
	
	\section{Nuclear Spin Bath Noise Spectrum}
	
	The presence of a 2D layer of proton nuclear spins on the diamond surface will produce a fluctuating magnetic field at the site of the surface spins $B^z(t)$ characterized by a magnetic power spectral density $V(\omega)$ where
	
	\begin{equation}
		B^z(t) = B^z_\perp(t) + B^z_\parallel(t).
	\end{equation}
	
	We define the z-axis as along the direction of the externally applied magnetic field, which we align along the NV axis.
	We calculate the projection of the magnetic field due to sum of nuclear spin magnetic moments along the direction of the applied external magnetic field.
	$B^z_\perp(t)$ is this projection due to components of the nuclear spin magnetic moment perpendicular to the direction of the external magnetic field, which rotate at the nuclear Larmor frequency, and $B^z_\parallel(t)$ is the sum of the contributions from the nuclear spin magnetic moments parallel to the direction of the external magnetic field.
	From the expression for $B^z(t)$, the noise spectrum $V(\omega)$ will have a low-frequency term due to slow time dependence in $B^z_\parallel(t)$ as well as higher frequency components due to the precession of the transverse magnetic moment, captured by $B^z_\perp(t)$.
	
	We can characterize the low frequency term in $V(\omega)$ following the treatment in~\cite{Taylor2008a}.  We model the flip-flop relaxation process of the proton nuclear spins as an exponentially correlated Gaussian fluctuating field $B^z_\parallel(t)$ having correlation function $\langle B^z_\parallel(t_0+t) B^z_\parallel(t_0)\rangle = \langle B_{\parallel,rms}^2 \rangle \exp(-t/\tau)$ where $\tau$ is the longitudinal correlation time of the nuclear spins and $B_{\parallel,rms}$ is the rms value of the magnetic field's longitudinal component projected along the z-axis.

	\subsection{Wiener-Khinchin Theorem}
	
	Using the correlation of the magnetic noise from the fluctuating nuclear spins, we construct the power spectrum for this noise using the Wiener-Khinchin Theorem.  The Wiener-Khinchin theorem states that given a random process $x(t)$ with autocorrelation function
	
	\begin{equation}
		C(t) = \langle x(t_0+t) x(t_0) \rangle,
	\end{equation}
	
	the spectral power density is the Fourier transform of the autocorrelation function
	
	\begin{equation}
		V(\omega) = \int_{-\infty}^\infty C(t) e^{-i\omega t} dt.
	\end{equation}
	
	For our calculations, we use the following definitions of Fourier transform and inverse Fourier transform:
	
	\begin{gather}
		X(\omega) = \int_{-\infty}^\infty x(t) e^{-i\omega t}dt\\ 
		x(t) = \frac{1}{2\pi}\int_{-\infty}^\infty X(\omega) e^{i\omega t}d\omega
	\end{gather}
	
	Using the Wiener-Khinchin theorem with the above definition of Fourier transform, we calculate the spectral power density for the low frequency noise associated with relaxation processes for proton nuclear spins as
	
	\begin{equation}
		V_1(\omega) = \int_{-\infty}^\infty \langle B_{\parallel,rms}^2 \rangle e^{-|t|/\tau} e^{-i\omega t} dt = 2 \langle B_{\parallel,rms}^2 \rangle \frac{\tau}{\omega^2\tau^2+1}.
	\end{equation}
	
	We apply the same procedure to determine the noise spectrum $V_2(\omega)$ due to proton nuclear spins precessing at their Larmor frequency $\omega_L$.  We begin by writing the correlation of the magnetic field noise:
	
	\begin{equation}
		\langle B_\perp(t_0+t)B_\perp(t_0) \rangle = \langle B_{\perp,rms}^2 \rangle e^{i\omega_L t - t/\tau}.
	\end{equation}
	
	We then apply the Wiener-Khinchin formula and take the Fourier transform of $\langle B_\perp(t_0+t)B_\perp(t_0) \rangle$ in order to arrive at the power spectral density
	
	\begin{equation}
		V_2(\omega) = \int_{-\infty}^\infty \langle B_{\perp,rms}^2 \rangle e^{i\omega_L |t|-|t|/\tau} e^{-i\omega t} dt = 2\langle B_{\perp,rms}^2 \rangle\frac{\tau}{(\omega-\omega_L)^2\tau^2+1}.
	\end{equation}
	
	We combine the contribution of the nuclear spin bath's component at zero frequency and at Larmor frequency to write the final nuclear spin noise power spectrum as
	
	\begin{equation}
		V(\omega) = 2\langle B_{\parallel,rms}^2 \rangle \frac{\tau}{\omega^2\tau^2+1} + 2\langle B_{\perp,rms}^2 \rangle\frac{\tau}{(\omega-\omega_L)^2\tau^2+1}.
	\end{equation}
	
	$\langle B_{\parallel,rms}^2 \rangle$ and $\langle B_{\perp,rms}^2 \rangle$ are geometrically related and we will next derive their relationship.  For all following calculations, we use CGS units.
	
	\subsection{Calculate $\langle B_{\parallel,rms}^2 \rangle$ from Longitudinal Component of the Nuclear Magnetic Moment}
	
	Here we wish to calculate the magnetic field at the site of surface spins from nuclear spins in the system.
	We follow the treatment presented in~\cite{Mamin2013}.
	An external field along the NV symmetry axis sets the longitudinal quantization axis $\mathbf{\hat{q}}$.  
	The system is comprised of shallow electronic spins near the surface of diamond with proton nuclear spins on the diamond surface.  
	All nuclear spins are unpolarized and produce a fluctuating magnetic field at the site of each surface spin, and we are interested in calculating the RMS amplitude of the component of this magnetic field along the quantization axis $\mathbf{\hat{q}}$ of the surface spins.  We begin by considering the longitudinal component of the magnetic moment of nuclear spins.  The component of magnetic field along $\mathbf{\hat{q}}$ from a single nuclear spin parallel or anti-parallel to $\mathbf{\hat{q}}$ is
	
	\begin{equation}
		B_q(\mathbf{r}) = \mathbf{\hat{q}}\cdot\mathbf{B} = \mathbf{\hat{q}}\cdot\frac{3\mathbf{\hat{r}}(\vec{m}_n\cdot\mathbf{\hat{r}})-\vec{m}_n}{r^3}=m_n[\frac{3(\mathbf{\hat{r}}\cdot\mathbf{\hat{q}})^2-1}{r^3}]
	\end{equation}
	
	where $\mathbf{r}$ is the vector from the nuclear spin to the surface electron spin and $m_n$ is the magnetic moment of the nuclear spin.  The mean square field from a collection of spins is
	
	\begin{equation}
		B^2_{\parallel,rms} = \langle[\sum_iB_q(\mathbf{r}_i)]^2\rangle.
	\end{equation}
	
	By expanding the sum, we get
	
	\begin{equation}
		B^2_{\parallel,rms} = \langle\sum_{i\ne j}B_q(\mathbf{r}_i) B_q(\mathbf{r}_j) + \sum_i B_q^2(\mathbf{r}_i)\rangle.
	\end{equation}
	
	Since the nuclear spins are random and uncorrelated, this reduces to
	
	\begin{equation}
		B^2_{\parallel,rms} = \sum_i B_q^2(\mathbf{r}_i).
	\end{equation}
	
	We can convert the sum to an integral by including the nuclear spin density $\rho_n$:
	
	\begin{equation}
		B^2_{\parallel,rms} = \rho_n \int_V[B_q(\mathbf{r})]^2 dV.
	\end{equation}
	
	We can define $\mathbf{\hat{q}}$ as the unit vector along the [1,1,1] axis and define $\mathbf{\hat{r}}$ in terms of its spherical coordinates:
	
	\begin{gather}
		\mathbf{\hat{q}} = [1,1,1]/\sqrt{3} \\
		\mathbf{\hat{r}} = [\sin(\theta)\cos(\phi),\sin(\theta)\sin(\phi),\cos(\theta)]
	\end{gather}
	
	where $\theta$ is the polar angle measured from the normal of the diamond surface, and $\phi$ is the aximuthal angle.  
	
	
	
	
	
	
	
	
	
	
	When the diamond is immersed in deutarated glycerol, the simplest assumption to make is that the proton nuclear spins form a 2D layer at the diamond surface~\cite{Staudacher2015,Loretz2014,DeVience2015}.
	Thus, for the case of surface spins which reside a depth d below the 2D layer of proton nuclear spins, we can relate the distance from the surface spins to a nuclear spin plane through the depth $d$ and polar angle $\theta$
	
	\begin{gather}
		\cos\theta = \frac{d}{r}\\
		\sin\theta = \sqrt{1-\frac{d^2}{r^2}}
	\end{gather}
	
	The rms magnetic field can be calculated using
	
	\begin{gather}
		B_{\parallel,rms}^2 = m_n^2\rho_n \int_0^{2\pi} \int_d^\infty [\frac{(\sqrt{1-\frac{d^2}{r^2}}\cos\phi+\sqrt{1-\frac{d^2}{r^2}}\sin\phi+d/r)^2-1}{r^3}]^2 r dr d\phi \\
		B_{\parallel,rms}^2 = m_n^2\rho_n\frac{3\pi}{8d^4}
	\end{gather}
	
	However, one must be careful when calculating the magnetic moment.  The magnetic moment is defined as
	
	\begin{equation}
		\vec{m} = g\mu_N\vec{S}
	\end{equation}
	
	where $g$ is the effective g-factor and $\mu_N$ is the nuclear magneton.  We can define
	
	\begin{gather}
		\langle m \rangle = g\mu_N\langle S\rangle \\
		m = g \mu_N \sqrt{S(S+1)} = \gamma_N\hbar\sqrt{S(S+1)}
	\end{gather}
	
	where $\gamma_N$ is the nuclear spin gyromagnetic ratio.
	
	
	
	For proton nuclear spins in the 2D plane of the diamod surface, the rms magnetic field on the surface spin z-axis from the longitudinal component of the magnetic field is given by
	
	\begin{equation}
		B_{\parallel,rms}^2 = \frac{3\pi}{8d^4}\rho_n(\gamma_N \hbar\sqrt{S(S+1)})^2
	\end{equation}
	
	for surface spins with depth d below the proton layer.  
	
	
	When we perform depth measurements, we place the diamond in undeuterated glycerol.
	For proton nuclear spins in the half space above the diamond surface, we can calculate the RMS magnetic field from the component of the magnetic moment paralel or antiparallel to $\mathbf{\hat{q}}$:
	
	\begin{gather}
		B_{rms}^2 = m_n^2 \rho_n \int_0^{\pi/2} \int_{d/\cos\theta}^{\infty} \int_0^{2\pi} [\frac{(\sin\theta\cos\phi+\sin\theta\sin\phi+\cos\theta)^2-1}{r^3}]^2 r^2 sin\theta d\phi dr d\theta
	\end{gather}
	
	where d is the depth of the surface spin.  Solving this integral yields
	
	\begin{gather}
		B_{rms}^2 = \frac{\pi m_n^2\rho_n}{8d^3}
	\end{gather}
	
	Using the expression for magnetic moment, this becomes
	
	\begin{equation}
		B_{rms}^2 = \frac{\pi\rho_n}{8d^3}(\gamma_N \hbar \sqrt{S(S+1)})^2
	\end{equation}
	
	\subsection{Calculate $\langle B_{\perp,rms}^2 \rangle$ from Transverse Component of the Nuclear Magnetic Moment}
	
	We follow the treatment in~\cite{Pham2016} to calculate $\langle B_{\perp,rms}^2 \rangle$ due to the transverse component of the nuclear spin magnetic moment.
	
	We can calculate the rms magnetic field from the transverse component of the magnetic moment of proton nuclear spins in the half space above the diamond surface:
	
	\begin{gather}
		B_{\perp,rms}^2 = \frac{2(g \mu_N \sqrt{S(S+1)})^2}{3}\rho \int_{-\infty}^{\infty} dx \int_{-\infty}^{\infty} dy \int_{d}^{\infty} dz \frac{(x+y+z)^2(r^2-xy-yz-xz)}{(x^2+y^2+z^2)^5}
	\end{gather}
	
	\begin{equation}
		B_{\perp,rms}^2 = \frac{5\pi\rho}{72 d^3}(g \mu_N \sqrt{S(S+1)})^2
	\end{equation}

	Using this technique, we can calculate the rms magnetic field from a 2D layer of proton spins on the diamond surface:
	
	\begin{gather}
		B_{\perp,rms}^2 =\frac{2 (g \mu_N \sqrt{S(S+1)})^2}{3}\rho \int_{-\infty}^{\infty} dx \int_{-\infty}^{\infty} dy \frac{(x+y+d)^2(r^2-xy-ya-xa)}{(x^2+y^2+d^2)^5}\\
		B_{\perp,rms}^2 = \frac{5\pi \rho}{24d^4}(g \mu_N \sqrt{S(S+1)})^2
	\end{gather}

	We find that the projection of the rms magnetic field from the longitudinal component of the magnetic moment on the surface spin z-axis is $3/\sqrt{5}$ times larger than the projection from the transverse component of the nuclear spin magnetic moment for both nuclear spins in the 2D plane as well as nuclear spins in the half space.
	
	\subsection{Noise Spectrum Equation}
	
	Using the relationship previously derived between the transverse and longitudinal $\langle B_{rms}^2 \rangle$ for the given geometry, we can write the expression for the nuclear spin bath noise spectrum as
	
	\begin{equation}
		V(\omega) = 2\langle B_{\parallel,rms}^2 \rangle (\frac{\tau}{\omega^2\tau^2+1} + +\frac{5}{9}\frac{\tau}{(\omega-\omega_L)^2\tau^2+1})
	\end{equation}
	
	where 
	
	\begin{equation}
		B_{\parallel,rms}^2 = \frac{3\pi}{8d^4}\rho_n(g\mu_N\sqrt{S(S+1)})^2
	\end{equation}
	
	for a system with proton nuclear spins residing on the 2D surface of the diamond and surface spins a depth d below.  We can make one further step and rewrite the power spectral density replacing $\langle B_{\parallel,rms}^2 \rangle$ with $W^2$, which are related by $\gamma_e^2$, the electron gyromagnetic ratio.  We recover the expression for the nuclear spin bath noise spectrum that we use in the main text:
	
	\begin{equation}
		V(\omega) = 2\frac{W^2}{\gamma_e^2} (\frac{\tau}{\omega^2\tau^2+1} + +\frac{5}{9}\frac{\tau}{(\omega-\omega_L)^2\tau^2+1}).
	\end{equation}
	
	\section{Characterizing NV Center - Surface Spin System}
	
	\subsection{NV Center Depth}
	
	In order to extract the depth of the NV centers used in this study, we placed the diamond in protonated glycerol and performed XY RF pulse sequences on the NV center to correlate the strength of the measured proton signal to the depth of the NV center.
	Following the treatment in~\cite{Pham2016}, when performing an XY-64 pulse sequence on one NV center, an NV center depth of d$_{NV}$ = (2.92 $\pm$ 0.24) nm was extracted.
	Depth measurements on other NV centers used in this study yielded similar results.
	
	\begin{figure}[h]
		\centering
		\includegraphics[width=0.5\textwidth]{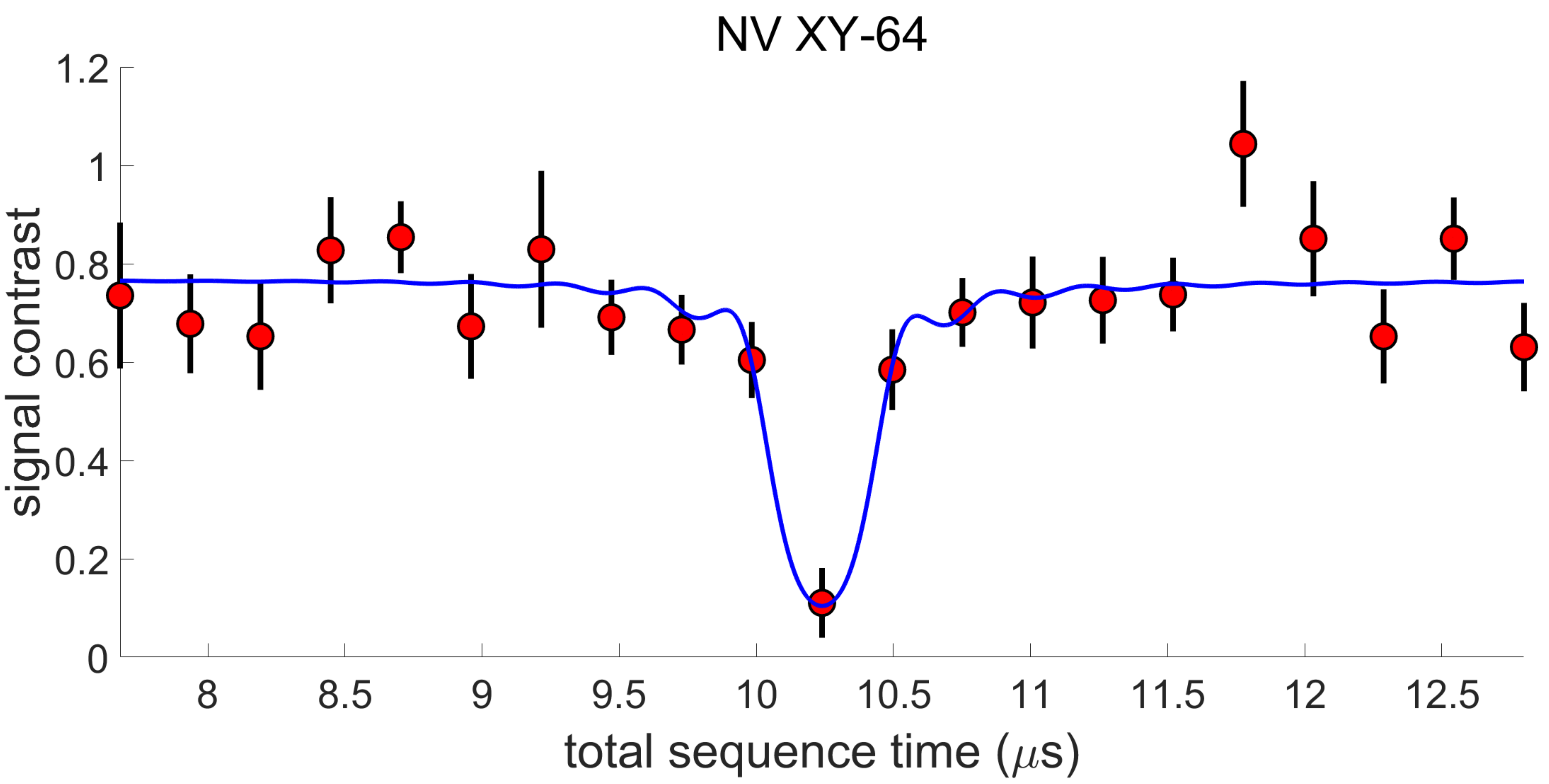}
		\caption{NV XY-64 measurement, with total sequence time centered around expected proton peak, performed in an external magnetic field of 735 G.  Fit (blue curve) to Eq. 1 in~\cite{Pham2016}, extracting a proton $B_{rms}$ corresponding to an NV center depth of d$_{NV}$ = (2.92 $\pm$ 0.24) nm.}
	\end{figure} 
	
	\subsection{The Double Electron-Electron Resonance (DEER) Sequence}
	
	The double elecron-electron resonance (DEER) sequence characterizes the strength of the dipolar interaction between the NV center and nearby surface electron spins.  
	The dipolar interaction Hamiltonian between the NV center $\mathbf{J}$ and surface spins $\mathbf{S_i}$ is written as
	
	\begin{equation}
		H_{NV-ss} = \sum_i \frac{\hbar^2 \gamma_e^2}{r^3_{NV-i}}[\mathbf{J}\cdot\mathbf{S_i} - 3\frac{(\mathbf{J}\cdot\mathbf{r_{NV-i}})(\mathbf{S_i}\cdot \mathbf{r_{NV-i})} }{r^2_{NV-i}}],
	\end{equation}
	
	where $\mathbf{r_{NV-i}}$ is the vector from the NV center to surface spin $i$.  When we align the external magnetic field along the NV center axis, we can make a secular approximation and only consider terms that commute with $S_i^z$ and $J^z$:
	
	\begin{equation}
		H_{NV-ss} = \sum_i \frac{\hbar^2 \gamma_e^2}{r^3_{NV-i}}(1-3\cos^2\theta_i)S_i^z J^z = \sum_i \hbar k_i S_i^z J^z
	\end{equation}
	
	where $\theta_i$ is the angle the vector $\mathbf{r_{NV-i}}$ makes with the external magnetic field, and the coupling strength is
	
	\begin{equation}
		k_i = \frac{\hbar \gamma_e^2}{r^3_{NV-i}}(1-3\cos^2\theta_i).
	\end{equation}
	
	The NV center $|m_s = 0\rangle$ state after performing the DEER sequence with total sequence time $t_{NV}$ is
	
	\begin{equation}
		S_{DEER} = \frac{1}{2}[1+\cos(\sum_i k_i S_i^z t_{NV})].
	\end{equation}
	
	After performing many averages with different initial surface spin projections, the final expected signal~\cite{Sushkov2014} is
	
	\begin{equation}
		S_{DEER} = \frac{1}{2}[1+\prod_i \cos(k_i t_{NV}/2)].
	\end{equation}
	
	For systems of strong coupling between the NV center and one surface electron spin, oscillations at the largest coupling strength are present at short time.  For the data presented in the main text fig. 1 and fig. 2, the NV spin-echo decays on a time scale $T_2^{NV}$ = (2.67 $\pm$ 0.13) $\mu$s and has a coupling strength of $k_{max}$ = (4.90 $\pm$ 0.57) $\mu$s$^{-1}$ to the strongest coupled surface spin.
	
	\subsection{The Correlation Spectroscopy Pulse Sequence}
	
	The correlation spectroscopy pulse sequence (fig. 2a) measures the autocorrelation of the surface spin $S_z$ state $\langle S_z(t_0+t)S_z(t_0) \rangle$~\cite{Laraoui2013}.  
	After the initial $\frac{\pi}{2}^{NV}_x - \pi^{NV}_y \pi^{SS} - \frac{\pi}{2}^{NV}_y$ pulses, the NV phase accumulation $\phi_1$ due to the surface spin $S_z$ state is projected along the NV z-axis.  
	If the NV state is read out at this time, the NV population would be $\sim \langle \sin \phi_1 \rangle$.  
	By applying two sequences of  $\frac{\pi}{2}^{NV}_x - \pi^{NV}_y \pi^{SS} - \frac{\pi}{2}^{NV}_y$ pulses, the initial and final surface spin $S_z$ projections are converted to population along the NV z-axis, with an amplitude related to $ \langle \sin \phi_1 \sin \phi_2 \rangle$ where $\phi_2$ is the phase the NV spin acquires during the final $\frac{\pi}{2}^{NV}_x - \pi^{NV}_y \pi^{SS} - \frac{\pi}{2}^{NV}_y$ pulses.
	
	For all correlation spectroscopy sequences that we perform, we normalize the result by the correlation spectroscopy contrast.
	We define the contrast as the NV spin population difference between when we run the correlation spectroscopy pulse sequence (fig. 2a) without any pulses applied during the grey shaded region and when we run the sequence with a pi pulse applied to the surface spins during this time.
	We normalize the final signal between +1 and -1, plotting 4$\langle S_z(t_0+t)S_z(t_0) \rangle$.
	
	\subsection{Surface Spin Pulse Sequences}
	
	In this section we present the pulse sequences that we use in the main text for fig. 2. All of the following sequences are applied during the shaded region of the correlation spectroscopy pulse sequence (fig. 2 (a)).
	
	\begin{figure}[h]
		\centering
		\includegraphics[width=0.6\textwidth]{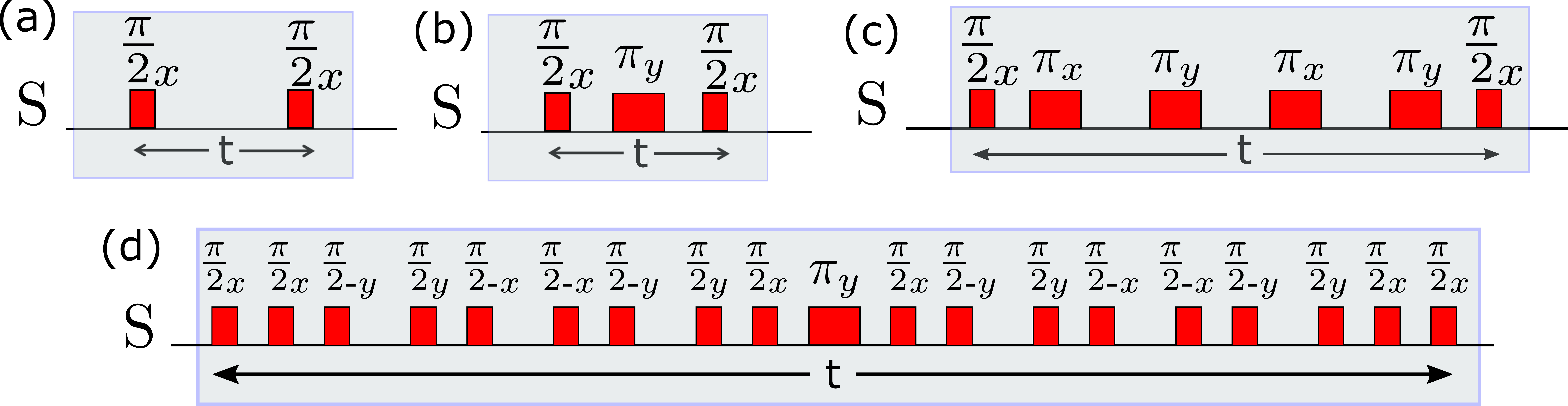}
		\caption{Pulse sequences applied to surface spins during the correlation spectroscopy pulse sequence. (a) Surface spin Ramsey, (b) surface spin echo, (c) surface spin XY-4, and (d) surface spin MREV-8 measurement, which we apply sandwiched in the free precession interval of the spin echo.}
	\end{figure}
	
	\subsection{Surface Spin Stability}
	
	Recent work has indicated that surface electron spins may not have stable positions but may be mobile on the diamond surface, potentially during the application of green laser light ~\cite{Dwyer2021}.  We directly probe the stability of the strongest coupled surface spin by measuring its state under the application of green laser light.  First, we polarize the surface electron spin by driving both the surface spins and the NV center at the same drive strength for a time set by the dipolar interaction between the NV center and the central surface spin.  We further calibrate the Hartmann-Hahn polarization transfer by observing maximal polarization transfer when the drive strength of the surface spins matches the drive strength of the NV center.
	
	\begin{figure}[h]
		\centering
		\includegraphics[width=0.8\textwidth]{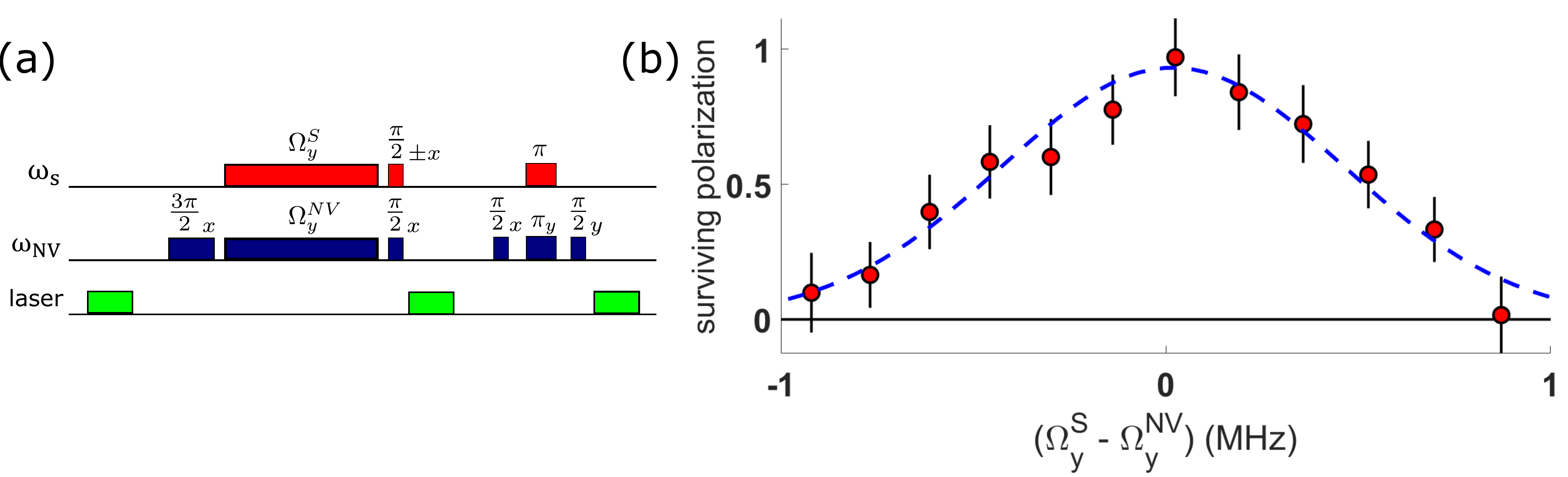}
		\caption{Hartmann-Hahn spin polarization transfer from the NV center to the strongest coupled surface spin. (a) HH polarization transfer sequence and readout of final NV center state contingent on polarization of central surface spin. (b) Measurement result of sweeping the surface spin drive where the resulting polarization of the central surface spin is read out using the NV center.}
	\end{figure}
	
	To study the stability of the strongest coupled surface spin, we sweep the duration of the green laser pulse after the polarization transfer.  The strongest coupled surface spin is stable under green laser light illumination for at least 50 $\mu$s.
	
	\begin{figure}[h]
		\centering
		\includegraphics[width=0.85\textwidth]{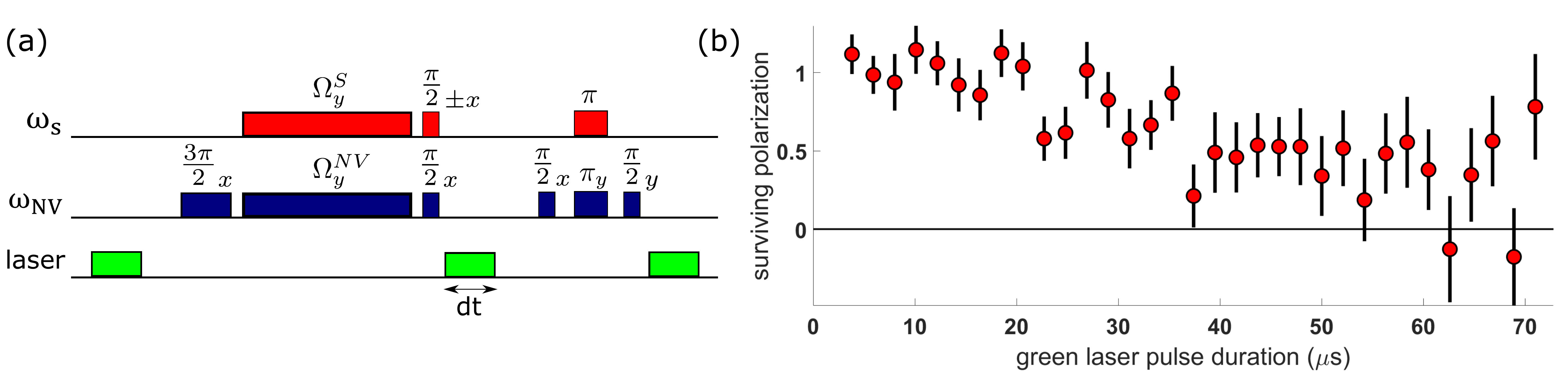}
		\caption{Surviving polarization of the central surface spin after green laser light illumination. (a) HH polarization transfer sequence and readout of final NV center state contingent on polarization of central surface spin while green NV repolarization laser pulse duration is swept. (b) Measurement result, normalized by the contrast of the $S_z$ autocorrelation measurement of equivalent initial time delay.}
	\end{figure}
	
	\section{Characterization of Surface Spin System}
	
	In this section, we extract parameters relevant to central surface spin dynamics including the strength and correlation time of the on-site disorder from the nuclear spin bath, the dipolar interaction strength between nearby surface spins, and the density of surface spins.
	
	\subsection{Decoherence due to the Nuclear Spin Bath}
	
	In a previous section, we derived the proton nuclear spin bath noise spectrum as a sum of Lorentzians with components centered at zero and at proton Larmor period:
	
	\begin{equation}
		V(\omega) = 2 \frac{W^2}{\gamma_e^2} (\frac{\tau}{\omega^2\tau^2+1}+\frac{5}{9}\frac{\tau}{(\omega-\omega_L)^2\tau^2+1})
	\end{equation}
	
	For the pulse sequences applied to the surface spins, we calculate the expected signal from the nuclear spin bath.  For a given sequence, the expected signal can be written as
	
	\begin{equation}
		S(t) = e^{-\chi(t)}
	\end{equation}
	
	where
	
	\begin{equation}
		\chi(t) = \gamma_e^2 \int \frac{d\omega}{2\pi}\frac{V(\omega)}{\omega^2}F(\omega t)
	\end{equation}
	
	where $F(\omega t)$ is the filter function of the pulse sequence and $t$ is the total sequence time.  In order to simplify calculations, we treat the peak at proton Larmor frequency $\omega_L$ as a delta function
	
	\begin{equation}
		V_2(\omega) = 2\pi \frac{5}{9} \frac{W^2}{\gamma_e^2} \delta(\omega-\omega_L).
	\end{equation}
	
	We select a delta function with a factor of 2$\pi$ in front so that the integral of the Lorentzian is matched by the integral of this delta function, and we numerically verify that for our given pulse sequences, the response due to a narrow Lorentzian is consistent with the response due to a delta function.  For the Ramsey pulse sequence with filter function $F_{Ramsey}(\omega t) = 2\sin^2(\omega t/2)$, the expected decay due to the nuclear spin bath is 
	
	\begin{equation}
		S_{Ramsey}(t) = e^{-(Wt/\sqrt{2})^2}e^{W^2 t^3/(6\tau)}e^{-2\frac{5 W^2}{9\omega_L^2}\sin^2(\omega_Lt/2)}.
	\end{equation}
	
	The filter function for the spin echo sequence is $F_{echo}(\omega t) = 8\sin^4(\omega t/4)$, and the expected signal decay due to the nuclear spin bath is
	
	\begin{equation}
		S_{echo}(t) =e^{-\frac{W^2 t^3}{12\tau}}e^{-\frac{40 W^2}{9 \omega_L^2}\sin^4(\omega_Lt/4)}.
	\end{equation}
	
	The XY-4 pulse sequence has the filter function $F_{XY-4}(\omega t) = 128\sin^6(\omega t/16)(\cos(3\omega t/16)+\cos(5\omega t/16))^2$ and the decay due to the nuclear spin bath is given by
	
	\begin{equation}
		S_{XY-4}(t) = e^{-\frac{13 W^2 t^3}{4500 \tau}}e^{-128\frac{5 W^2}{9 \omega_L^2}\sin^6(\omega_Lt/16)(\cos(3\omega_Lt/16)+\cos(5\omega_Lt/16))^2}.
	\end{equation}
	
	The MREV-8 sequence in an echo has a filter function given by
	
	\begin{equation}
		F_{MREV}(\omega t) = 16(1+2\cos(\omega t/12))^2(\sin(\omega t/12)-\sin(\omega t/6)^4(3-4\cos(\omega t/24)+3\cos(\omega t/12)-2\cos(\omega t/8)+\cos(\omega t/6))
	\end{equation}
	
	with a resulting signal of 
	
	\begin{equation}
		\begin{split}
			-\log(S_{MREV}(t)) = \frac{49 W^2 t^3}{2592\tau}+16\frac{5 W^2}{9 \omega_L^2}(1+2\cos(\omega_Lt/12))^2(\sin(\omega_Lt/12)-\sin(\omega_Lt/6)^4(3-4\cos(\omega_Lt/24)+\\
			3\cos(\omega_Lt/12)-2\cos(\omega_Lt/8)+\cos(\omega_Lt/6)).
		\end{split}
	\end{equation}
	
	For each of the pulse sequences, the resulting signal has two components: a decay envelope due to the noise spectrum peak at zero frequency and a modulation due to the signal at proton Larmor frequency.
	
	\subsection{Decay due to Dipolar Interactions}
	
	While the central surface spin Ramsey, echo, and XY measurements will be affected by the nuclear spin bath, the dipolar interations between the central surface spin and nearby surface spins will also cause decoherence.  If there were one other surface spin in the system interacting with the central surface spin with dipolar interaction strength $J_1= \frac{\hbar\gamma_e^2}{r^3}(1-3\cos^2\theta)$, the resulting signal for the Ramsey, echo, and XY sequences would be
	
	\begin{equation}
		S_{Ramsey,XY-4,Echo}(t) = \frac{1}{2}[\cos(3J_1 t/4)+\cos(J_1 t/4)].
	\end{equation}
	
	On short time scales, the central surface spin is mostly sensitive to nearest neighbor dipolar interactions, so we approximate the decay profile as a stretched exponential with a power of 2, $e^{-(t/T_2)^2}$, with decay time $T_2 \approx 1/J_1$, which we numerically verify.
	If we assume that the decays due to dipolar interactions and the nuclear spin bath are independent processes, then we can express the overall Ramsey, echo, and XY-4 decays as the product of the stretched exponential decay due to dipolar interactions with the decay and modulation due to the nuclear spin bath.
	Therefore, as an example, the fit function for the XY-4 measurement becomes
	
	\begin{equation}
		S_{XY-4}(t) = \exp(-(t/T_2)^2)\exp(-\frac{13 W^2 t^3}{4500 \tau})\exp(-128\frac{5 W^2}{9 \omega_L^2}\sin^6(\omega_L t/16)[\cos(3\omega_L t/16)+\cos(5\omega_L t/16)]^2)
	\end{equation}
	
	where $\omega_L$ is the proton Larmor frequency.  We can show that the dominant source of decoherence in the echo and XY-4 sequences comes from dipolar interactions between surface spins by comparing the decay rate of the echo, XY-4, and MREV-8 measurements.  If the nuclear spin bath were the primary source of decoherence for the echo or XY-4 sequences, there would be an extension of the coherence time between the echo and XY-4 measurement results.  However, the echo and XY-4 measurements decay over the same timescale.  Moreover, after applying an MREV-8 sequence, which decouples surface spins from homonuclear dipolar interactions, the coherence time increases, suggesting that dipolar interactions between surface spins are the primary contributor to decoherence in echo and XY-4 measurements.
	
	\begin{figure}[h]
		\centering
		\includegraphics[width=0.5\textwidth]{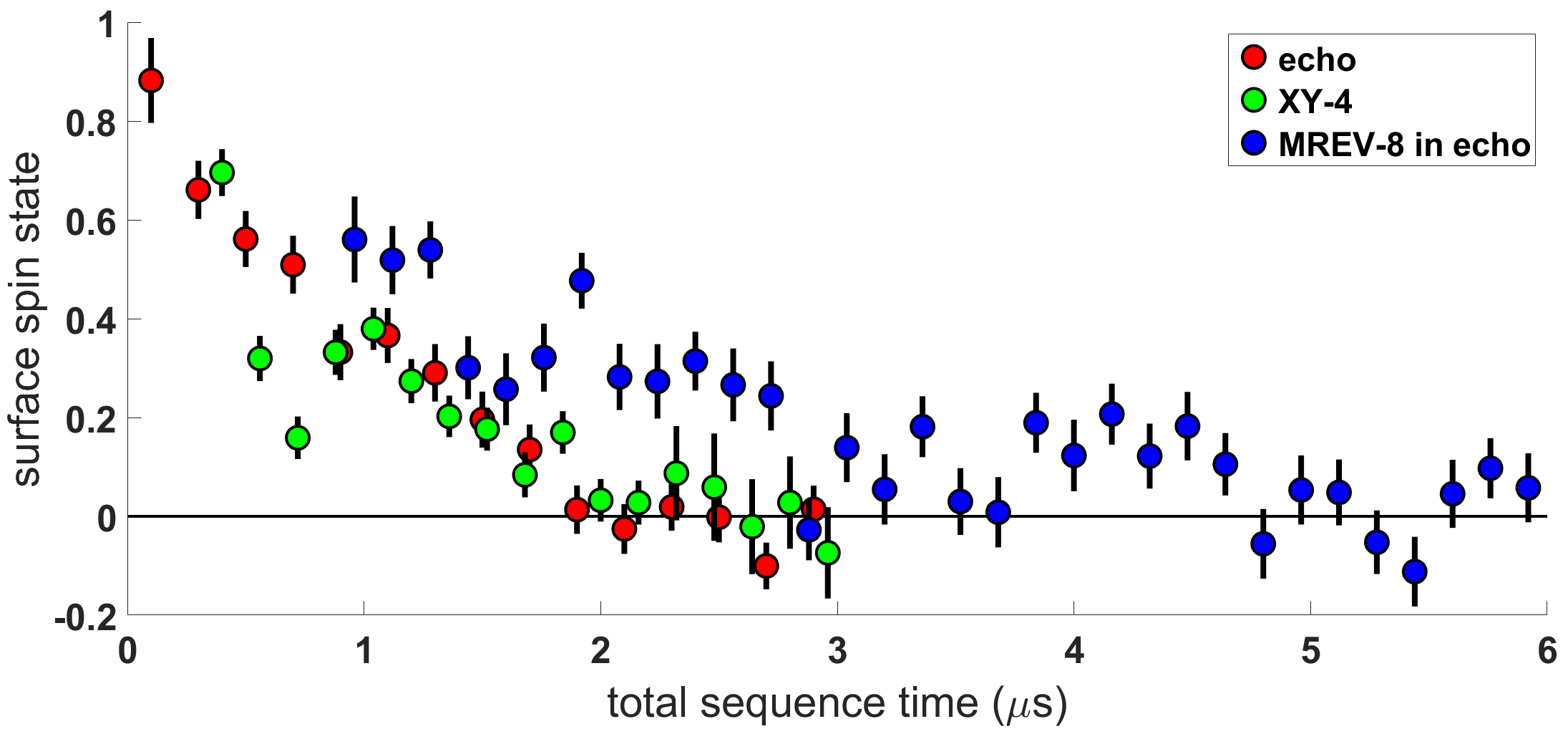}
		\caption{Comparison of surface spin echo, XY-4, and MREV-8 measurements, showing extension of coherence time with MREV-8.}
	\end{figure}

	\subsection{Extraction of Surface Spin On-site Disorder Strength W}
	
	In the previous sections, we calculated the expected signals given dipolar interactions and the nuclear spin bath.  For the experimentally relevant regime where the nuclear spin correlation time $\tau$ is much longer than the sequence duration, a detuned Ramsey sequence decay is given by
	
	\begin{equation}
		S_{Ramsey}(t) = \cos(\Delta t)e^{-(t/T_2)^2}e^{-(W t/\sqrt{2})^2}e^{-2\frac{5 W^2}{9 \omega_L^2}\sin^2(\omega_Lt/2)}
	\end{equation}
	
	where $\Delta$ is the detuning between the surface spin splitting $\omega_S$ and the microwave drive frequency.  By fitting the echo and XY-4 measurements, $T_2$ can be extracted, so $W$ is the only free parameter in a Ramsey fit.

	\subsection{Extraction of Proton Nuclear Spin Signal Correlation Time $\tau$}
	
	In a previous section, we found the MREV-8 decay to be given by
	
	\begin{equation}
		\begin{split}
			-\log(S_{MREV}(t)) = \frac{49 W^2 t^3}{2592\tau}+16\frac{5 W^2}{9 \omega_L^2}(1+2\cos(\omega_Lt/12))^2(\sin(\omega_Lt/12)-\sin(\omega_Lt/6)^4(3-4\cos(\omega_Lt/24)+\\
			3\cos(\omega_Lt/12)-2\cos(\omega_Lt/8)+\cos(\omega_Lt/6)).
		\end{split}
	\end{equation}
	
	Since MREV-8 decouples surface spins from dipolar interactions with other surface spins, the decay envelope is determined by the disorder width W and the nuclear spin correlation time $\tau$.  We extracted the disorder width from the Ramsey decay so by fitting the MREV-8 decay, we extract the nuclear spin correlation time $\tau$.
	
	\subsection{Extraction of Surface Spin Density}
	
	We extract the average surface spin density by comparing the results of numeric simulations to averaged experimental data.  We begin by averaging together all XY-4 data taken at an external magnetic field of 730 G and fit the periodic features in the data, extracting a value of W of 4.13 $\pm$ 0.24 $\mu$s$^{-1}$.  Next we perform numeric simulations with a central surface spin and five neighboring surface spins.  We perform simulations for different values of average separation between spins and average together the results for 500 averages with random spin positions for each value of average separation.  We extract an average surface spin separation of 8.4 $\pm$ 1.3 nm.  This corresponds to an average interaction strength J = $J_0n^{3/2}$ = (0.57 $\pm$ 0.25) $\mu$s$^{-1}$.
	
	\begin{figure}[h]
		\centering
		\includegraphics[width= .8\textwidth]{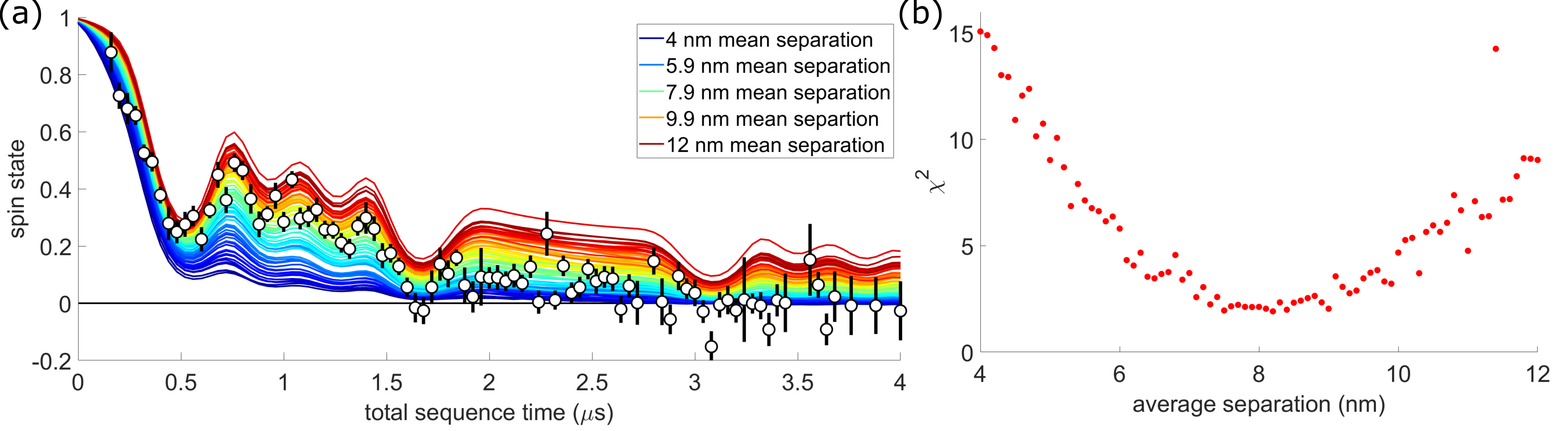}
		\label{fig.Density}
		\caption{(a) Simulation results with averaged experimental data (white dots). (b) $\chi^2$ vs average separation, where an average separation between surface spins of 8.4 $\pm$ 1.3 nm is extracted.}
	\end{figure}
	
	\subsection{Effects of Finite Pulse Duration}
	
	In all measurements, pulses are applied with a 12.5 MHz drive strength (40 ns $\pi$ pulse).
	The total drive duration for the Ramsey (0.04 $\mu$s), echo (0.08 $\mu$s), and XY-4 (0.2 $\mu$s) sequences are significantly shorter than the timescale of the decay, but for the MREV-8 sequence (0.4 $\mu$s), we consider the effects of the finite duration of the drive on the decay.  
	We do this by comparing the numeric simulation results of the MREV-8 pulse sequence with a nuclear spin bath for the case of finite pulse duration and infinitely fast pulses.  
	We plot on the x-axis the total evolution time, which includes the evolution time during the pulses for the case of finite pulse duration.  
	We can account for the finite pulse duration by adding the time of the drive to the total free precession time.
	
	\begin{figure}[h]
		\centering
		\includegraphics[width= .5\textwidth]{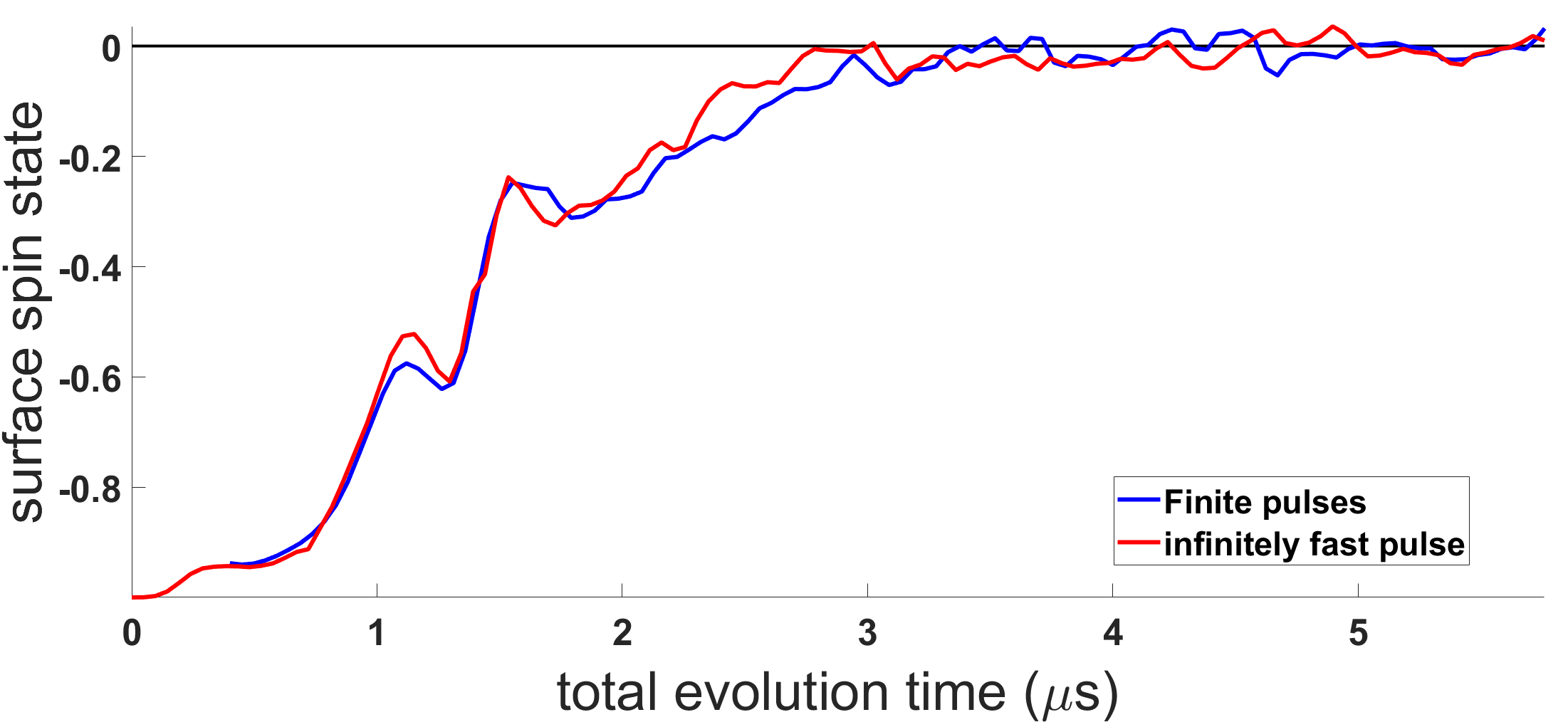}
		\caption{Simulation results for finite pulses versus infinitely fast pulses.  For infinitely fast pulses, the x-axis is given by the total free precession time and for finite pulses, the x-axis corresponds to the free precession time between pulses plus the finite duration of the pulses.}
		\label{fig.HHPolT2star}
	\end{figure}
	
	\subsection{NV Center's Effects on Surface Spin Dynamics}
	
	With the ability polarize the central surface spin using the NV center, we can study the effect the NV center spin state has on surface spin dynamics.  We can achieve this by first polarizing the central surface spin and then preparing the NV center in a particular spin state.  If the NV center is in the m$_s$ = 0 state, it will have no effect on the surface spins through the dipolar interaction.  However, if the NV center is in either the m$_s$ = 1 or the m$_s$ = -1 state, it will affect the surface spins through the dipolar interaction.  We can perform a surface spin Ramsey experiment after polarizing one surface spin with the NV in either m$_s$ = 0 or m$_s$ = -1 to observe the effect of the dipolar interaction of the NV center on the central surface spin.  Note that the drive pulses to the surface spins are detuned, which produce the oscillations.

	\begin{figure}[h]
		\centering
		\includegraphics[width= .6\textwidth]{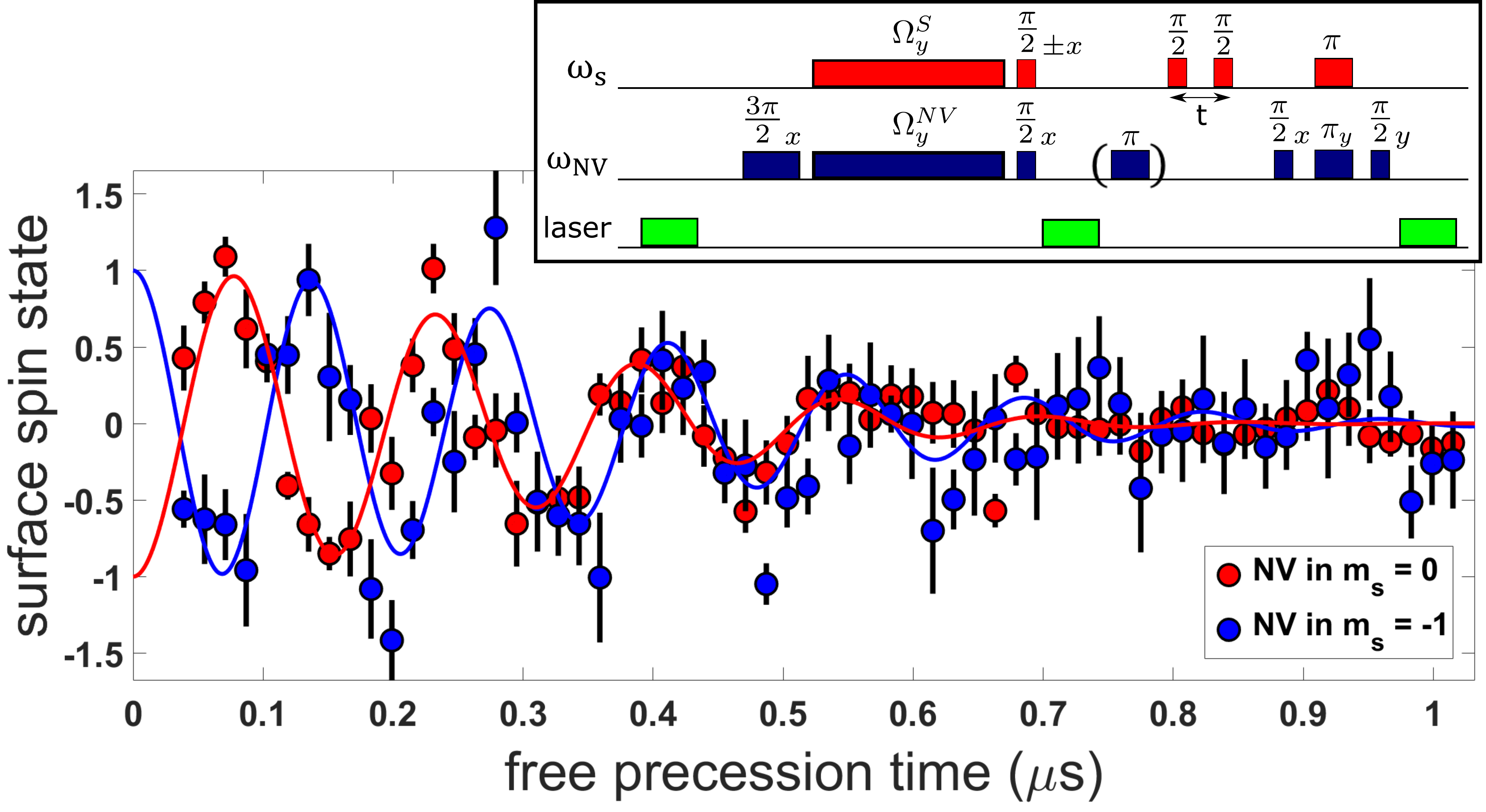}
		\caption{Surface spin detuned Ramsey measurement with the NV center in m$_s$ = 0 or m$_s$ = -1.  The oscillations are due to detuned surface spin pulses, and we observe a difference in oscillation frequency between the red and blue curves equal to the dipolar interaction between the NV center and the central surface spin.  Inset: pulse sequence with Hartmann-Hahn polarization transfer followed by Ramsey sequence applied to surface spins and then NV center state readout pulses.}
		\label{fig.HHPolT2star}
	\end{figure}
	
	The difference in oscillation frequency is given by the dipolar interaction strength between the NV center and the central surface spin and acts to further detune the surface spins when the NV is in m$_s$ = -1. 
	
	We can further observe the effect of the NV center spin state on surface spin dynamics by measuring the surface spin T$_z$ of the strongest coupled surface spin after polarization with the NV in either m$_s$ = 0 or m$_s$ = -1.
	
	\begin{figure}[h]
		\centering
		\includegraphics[width= .6\textwidth]{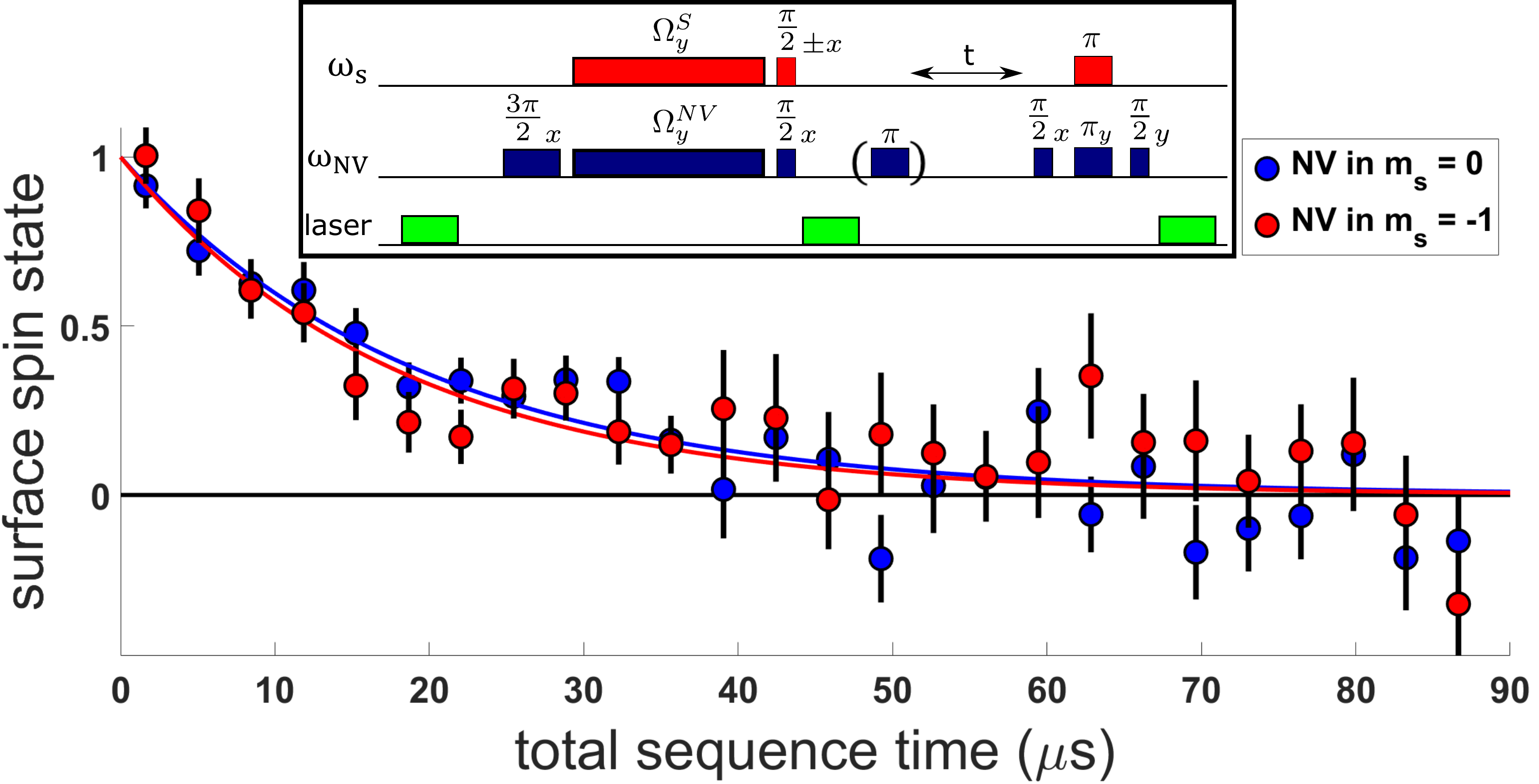}
		\caption{Surface spin T$_z$ measurement after polarization with the NV center in either m$_s$ = 0 or m$_s$ = -1. Measurement results indicate no difference in relaxation rate between the two cases, to within measurement uncertainty.  Inset: pulse sequence with Hartmann-Hahn spin polarization following by a swept delay time before NV pulses read out the final surface spin state.}
		\label{fig.HHPolT1}
	\end{figure}
	
	We observe that the dipolar interaction between the NV center and the surface spins does not affect the central surface spin relaxation.  For our parameter ranges of NV - surface spin dipolar interaction strength, surface spin - surface spin dipolar intreaction strength, and on-site nuclear spin disorder width, we can numerically verify that the relaxation of the central surface spin is dominated by the effects of the nuclear spin disorder and surface spin - surface spin dipolar interactions.  Thus, we can neglect the effect of the NV center in our resonance counting treatment.
	
	\section{Resonance Counting Theory}
	
	In this section, we present the single-particle resonance counting theory for the case of time-varying on-site fields at the sites of the surface spins.  We follow the treatment presented in~\cite{Kucsko2018}, which also treats the case of quasi-static on-site fields.
	
	\subsection{Dynamic Disorder Model}
	
	We estimate the survival probability of a central spin based on simple resonance counting arguments.
	The probability of a given spin being on resonance with the central spin is given by $\frac{\beta J_0/r^3}{W}$, where $J_0 = \hbar \gamma_e^2$, $J_0/r^3$ is the interaction strength of the central spin with a bath spin a distance $r$ away, $W$ is the disorder width, and $\beta$ is a dimensionless parameter that characterizes the resonance probability, of order unity.
	We consider the case of time dependent disorder, where the value of on-site detuning $\delta_j$ changes over time by uncorrelated jumps at a rate $\gamma = 1/\tau$.
	We consider the case where $W \geq nJ_0 > 1/\tau$.
	As presented in~\cite{Kucsko2018}, two spins $j$ and $i$ are on resonance at time t if (1) $|\delta_j(t)-\delta_i(t)|<\beta J_0/r^3$ at any time less than t (main text fig. 3(e)) and (2) the interaction occurs on a time-scale $t>\frac{1}{\beta J_0/r^3}$.
	
	The probability of having a resonant pair is given by
	
	\begin{equation}
		P_{res}(t) = 1-e^{-\frac{\beta J_0/r^3}{W}\frac{t}{\tau}}(1-\frac{\beta J_0/r^3}{W}).
	\end{equation}
	
	For a 2D spin system with density $n$, we can calculate the probability of a central spin having no resonance as
	
	\begin{gather}
		P(t) = \exp(-\int_{r_0}^{R(t)} 2 \pi n r P_{res}(t) dr).
	\end{gather}
	
	Note that $R(t) = (J_0 t)^{1/3}$.  When $\frac{\beta J_0/r^3}{W}$ is small (at large times), $(1-\frac{\beta J_0/r^3}{W})$ can be rewritten as $\exp(-\frac{\beta J_0/r^3}{W})$ and 
	
	\begin{equation}
		P_{res}(t) = 1 - e^{-\frac{\beta J_0/r^3}{W}(\frac{t}{\tau}+1)}.
	\end{equation}
	
	We can write the probability that the central spin has no resonance as 
	
	\begin{equation}
		P(t) = \exp(-\int_{r_0}^{R(t)} 2\pi n r (1 - e^{-\frac{\beta J_0/r^3}{W}(\frac{t}{\tau}+1)}) dr).
	\end{equation}
	
	To simplify the expression, we can take the $-log$ of each side:
	
	\begin{equation}
		-\log(P(t)) = \int_{r_0}^{R(t)} 2\pi n r (1 - e^{-\frac{\beta J_0/r^3}{W}(\frac{t}{\tau}+1)} dr).
	\end{equation}
	
	We can next perform a change of variables to evaluate the integral.  We define
	
	\begin{equation}
		z = \frac{\beta J_0/r^3}{W}(\frac{t}{\tau}+1).
	\end{equation}
	
	We can rewrite this in terms of r as
	
	\begin{equation}
		r = (\frac{\beta J_0/z}{W}(\frac{t}{\tau}+1))^{1/3}.
	\end{equation}
	
	Using the fact that $2 r dr = d(r^2)$, we can rewrite the expression for $P(t)$ in terms of z:
	
	\begin{equation}
		-\log(P(t)) = \int_{r(z,t)}^{R(z,t)} n \pi (1 - e^{-z}) d(\frac{\beta J_0/z}{W}(\frac{t}{\tau}+1))^{2/3}
	\end{equation}
	
	which becomes
	
	\begin{equation}
		-\log(P(t)) = \int_{z_0(r_0,t)}^{z(r,t)} n \pi (1 - e^{-z})(-2/3)z^{-5/3} dz (\frac{\beta J_0}{W}(\frac{t}{\tau}+1))^{2/3}.
	\end{equation}
	
	We can then rewrite $-\log(P(t))$ as the product of two terms:
	
	\begin{gather}
		-\log(P(t)) = P_1(t) P_2(t)\\
		P_1(t) = \int_{z_0(r_0,t)}^{z(r,t)} \pi (1 - e^{-z})(-2/3) z^{-5/3} dz\\
		P_2(t) = (n^{3/2}\frac{\beta J_0}{W}(\frac{t}{\tau}+1))^{2/3}
	\end{gather}

	In the limit $t \gg \tau$, $P_2(t)$ simplifies to
	
	\begin{equation}
		P_2(t) = (\frac{n^{3/2}\beta J_0}{W}\frac{t}{\tau})^{2/3}.
	\end{equation}
	
	We want to understand the time dependence in $P_1(t)$ so we define
	
	\begin{gather}
		z_0(r_0,t) = \frac{\beta J_0/r_0^3}{W}(t/\tau + 1)\\
		z(R,t) = \frac{\beta J_0/R(t)^3}{W}(t/\tau + 1)
	\end{gather}
	
	Since $R(t) = (J_0 t)^{1/3}$, in the limit $t \gg \tau$, z has weak dependence on time.  We can numerically integrate $P_1(t)$ in order to understand the time dependence, and we plot $P_1(t)$ and $P_2(t)$ for comparison.
	
	\begin{figure}[h]
		\centering
		\includegraphics[width= .5\textwidth]{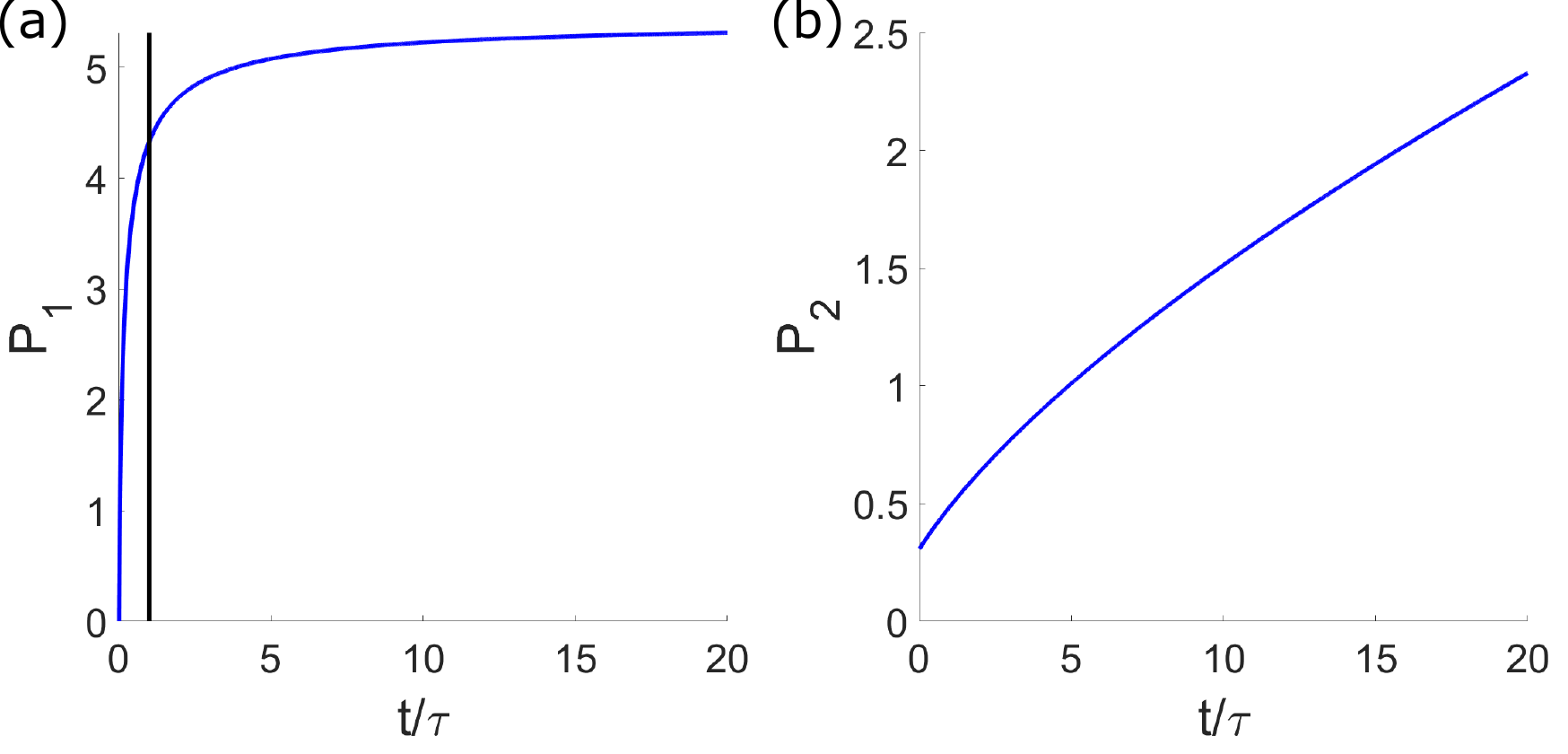}
		\caption{Plots of (a) $P_1(t)$ and (b) $P_2(t)$ with $r_0$ = 2 nm, $\beta$ = 1, $\tau$ = 15 $\mu$s, $W$ =  3.77 $\mu$s$^{-1}$, and $J_0$ = 326.7 nm$^3$/$\mu$s. $P_2(t)$ has a stronger dependence on time than $P_1(t)$ for times t$>\tau$.}
		\label{fig:fig1}
	\end{figure} 
	
	For times $t>\tau$, there is some dependence on time in $P_1(t)$, but it is not as strong as the time dependence in $P_2(t)$.
	Therefore, we can write the probability of having a resonant pair as 
	
	\begin{equation}
		P(t) = \exp(-\alpha (\frac{n^{3/2}\beta J_0}{W\tau}(t+\tau))^{2/3})
	\end{equation}
	
	where $\alpha$ is the average value of $P_1$ at times t$>\tau$.  
	For our parameter range, we set $\alpha = 5$, and we have numerically verified that $P(t)$ in Eq. E16 is consistent with Eq. E4 numerically integrated for relevant parameters.  
	From this expression, we can see that changing $W$ or $\tau$ simply amounts to changing the timescale of the decay.
	For times much longer than $\tau$, we can further simplify $P(t)$ to
	
	\begin{equation}
		P(t) = \exp(-\alpha (\frac{n^{3/2}\beta J_0}{W\tau}t)^{2/3}).
	\end{equation}
	
	
	
	\subsection{Two Sources of Dynamic Disorder}
	
	In this section we discuss the effects of having 2 sources of dynamic disorder with disorder widths $W_1$ and $W_2$ and correlation times $\tau_1$ and $\tau_2$.  We define an effective disorder width $W_e$ by adding the disorder widths from the 2 sources in quadrature:
	
	\begin{equation}
		W_e = \sqrt{W_1^2+W_2^2}.
	\end{equation}
	
	We can determine the effective correlation time by adding the correlation rates of the two disorder sources, weighted by the strength of each disorder source, as follows
	
	\begin{equation}
		\sqrt{1/\tau_e} = \frac{W_1\sqrt{1/\tau_1}+W_2\sqrt{1/\tau_2}}{W_e}.
	\end{equation}
	
	We have numerically verified these relationships.
	For the central surface spin system, the nuclear spin disorder has width W and correlation time $\tau$ while the positional/interaction disorder has width $J=1/T_2$ and correlation time $T_z$.
	
	
	
	\section{S$_z$ Autocorrelation Scaling}
	
	For the 7 systems with strong coupling between the NV center and a central surface spin presented in Fig. 3 of the main text, we have independently measured and extracted T$_z$, T$_2$, $\tau$, and W, and present those values below.
	
	\begin{figure}[h]
		\centering
		\includegraphics[width= .7\textwidth]{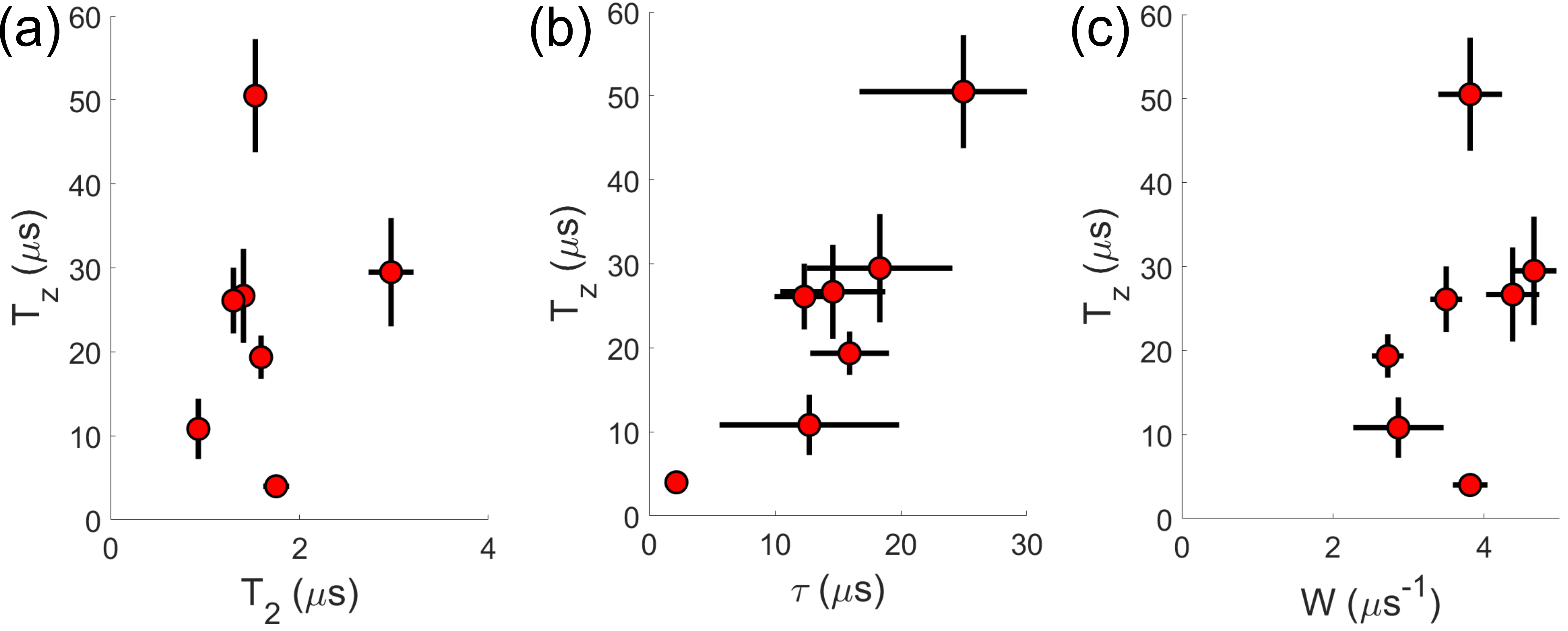}
		\caption{Dependence of T$_z$ on different system parameters.  (a) T$_z$ vs T$_2$, (b) T$_z$ vs $\tau$, and (c) T$_z$ vs W}
		\label{fig:fig1}
	\end{figure}

	\section{Spin-lock Driving Measurements (T$_{1\rho}$)}
	
	In order to control the strength of the nuclear spin bath disorder, we perform spin-lock driving measurements on the surface spin system.
	In this section we will discuss our treatment and understanding of those results.
	We begin by considering the Hamiltonian in the frame rotating with  respect to the bare splitting $\omega_S = \gamma_e B^z_{ext}$.
	The resulting Hamiltonian for the central surface spin $S_j$ from the main text plus an additional driving term in the rotating fame is
	
	\begin{equation}
		H_j' = \hbar \gamma_e B^z_j(t) S_j^z +\Omega S_j^y + \sum_i \frac{\hbar^2 \gamma_e^2}{r_{ij}^3}(1-3\cos^2\theta_{ij})[S_i^z S_j^z - \frac{1}{4}(S_i^+ S_j^- + S_i^- S_j^+)]
	\end{equation}
	
	where $\Omega$ is the driving strength.
	
	We consider the case of a two level system where spin operators are given by $S^\alpha = \frac{1}{2}\sigma^\alpha$.  One may further move into a second interaction picture with respect to $H_1 = \Omega S^y = \frac{1}{2} \Omega \sigma^y$.  Our spin operators are then transformed by replacing $S^\alpha$ with $U^\dagger(t) S^\alpha U(t)$, where $U(t) = \exp(-i \Omega t \sigma^y/2)$.  Our initial spin operators are transformed as follows:
	
	\begin{gather}
		S^x(t) \rightarrow \cos(\Omega t)S^{x\prime} + \sin(\Omega t)S^{z\prime}\\
		S^y(t) \rightarrow S^{y\prime}\\
		S^z(t) \rightarrow \cos(\Omega t)S^{z\prime} - \sin(\Omega t)S^{x\prime}
	\end{gather}
	
	where $S^{\alpha\prime}$ is the spin operator in the doubly rotated frame.  The resulting dipolar Hamiltonian has half the interaction strength as in the singly rotating frame:
	
	\begin{equation}
		H_{j}^{dd} = \sum_{ij} \frac{\hbar^2\gamma_e^2}{r_{ij}^3}(1-3\cos^2\theta_{ij})[\frac{1}{4} S_i^{z\prime} S_j^{z\prime} + \frac{1}{4}S_i^{x\prime} S_j^{x\prime} - \frac{1}{2} S_i^{y\prime} S_j^{y\prime}].
	\end{equation}
	
	In the doubly rotating frame with respect to the drive, the Hamiltonian due to the on-site fields from the nuclear spin bath becomes
	
	\begin{equation}
		H_j^{nuc} = \hbar \gamma_e B^z_{j}(t) [\cos(\Omega t)S_j^{z\prime} - \sin(\Omega t)S_j^{x\prime}].
	\end{equation}
	
	In this new basis, the energy eigenstates are split by the effective Rabi frequency of the drive $\Omega_{eff} = \sqrt{\Omega^2+\delta^2}$ where $\delta$ is the on-site detuning value, characterized by the power spectral density $V(\omega)$.
	Following the treatment in ~\cite{Kucsko2018}, the effective detuning width at strong driving strength scales as $W_{eff} = \frac{W^2}{\sqrt{2}\Omega}$.  
	We can numerically calculate $\delta$ values from the power spectral density $V(\omega)$ and calculate $\Omega_{eff}$ for different values of drive strength $\Omega$, verifying that at large drive strengths, the effective on-site disorder width scales as $W_{eff} = \frac{W^2}{\sqrt{2}\Omega}$.
	
	
	For weaker drives, the presence of nuclear spin bath low frequency noise will directly drive transitions between the dressed states of the surface spins, split by drive strength $\Omega$.  We can treat the nuclear spin bath noise spectrum as a perturbation and directly calculate the transition probability and relaxation rate given $V(\omega)$.  First, let's assume that the nuclear spin bath noise is monochromatic at frequency $\omega$:
	
	\begin{equation}
		H' = \hbar \gamma_e B_0 S^z \cos(\omega t).
	\end{equation}
	
	The matrix element $Q_{ab}$ is given by 
	
	\begin{equation}
		Q_{ab} = -\frac{1}{2} \hbar \gamma_e B_0.
	\end{equation}
	
	We can then calculate the transition probability as
	
	\begin{gather}
		P_{ab}(t) = \frac{|Q_{ab}|^2}{\hbar^2}\frac{\sin^2[(\Omega-\omega)t/2]}{(\Omega-\omega)^2}\\
		P_{ab}(t) = \frac{\hbar^2\gamma_e^2 B_0^2/4}{\hbar^2}\frac{\sin^2[(\Omega-\omega)t/2]}{(\Omega-\omega)^2}.
	\end{gather}
	
	The above expressions are all for monochromatic perturbations.  In our system, $B(t)$ can be described from a power spectrum $V(\omega)$, so we need to take this into account.  The energy density in CGS units of an electromagnetic wave is $U = \frac{B_0^2}{8\pi}$.  We can then rewrite the transition probability as
	
	\begin{equation}
		P_{ab}(t) = 2\pi U \gamma_e^2 \frac{\sin^2[(\Omega-\omega)t/2]}{(\Omega-\omega)^2}.
	\end{equation}
	
	Since the central surface spin is exposed to magnetic noise at a range of frequencies, we rewrite U as $\rho(\omega)d\omega$ where $\rho(\omega)d\omega$ is the energy density in the frequency range $d\omega$, and we can rewrite the transition probability as
	
	\begin{equation}
		P_{ab} = 2\pi\gamma_e^2 \int_0^\infty \rho(\omega)\frac{\sin^2[(\Omega-\omega)t/2]}{(\Omega-\omega)^2}d\omega.
	\end{equation}
	
	Since $\frac{\sin^2[(\Omega-\omega)t/2]}{(\Omega-\omega)^2}$ is sharply peaked around $\Omega$, we can replace $\rho(\omega)$ by $\rho(\Omega)$ and take it outside of the integral:
	
	\begin{equation}
		P_{ab} = 2\pi\gamma_e^2 \rho(\Omega) \int_0^\infty \frac{\sin^2[(\Omega-\omega)t/2]}{(\Omega-\omega)^2}d\omega.
	\end{equation}
	
	We can solve the integral by noting that
	
	\begin{equation}
		\int_{-\infty}^{\infty} \frac{\sin^2x}{x^2}dx = \pi.
	\end{equation}
	
	We find that
	
	\begin{equation}
		P_{ab} = 2\pi\gamma_e^2 \rho(\Omega)t\pi/2.
	\end{equation}
	
	We find that the transition rate ($R = dP/dt$) is
	
	\begin{equation}
		R_{ab} = \pi^2\gamma_e^2\rho(\Omega)
	\end{equation}
	
	where $\rho(\Omega)$ is the energy in the field at $\Omega$ and directly corresponds to the value of the power spectral density $S(\omega)$ at that frequency.
	
	\begin{figure}[h]
		\centering
		\includegraphics[width= .95\textwidth]{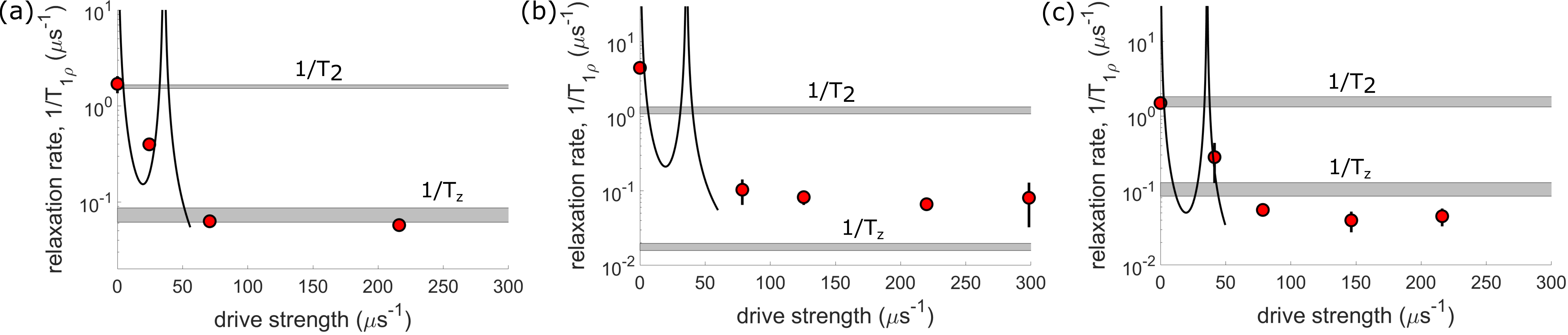}
		\caption{T$_{1\rho}$ relaxation rate as a function of drive strength for the 3 other NV center - surface spin system data sets shown in the main text fig 4c.  Black line represents the predicted relaxation rate from extracting W from each measurement's Ramsey decay and estimating $\tau$.  1/T$_2$ and 1/T$_z$ decay rates shown by the gray bands, independently extracted for each central surface spin system.}
		\label{fig:T1rho2}
	\end{figure} 

\end{adjustwidth}

\bibliography{library}

\begin{thebibliography}{10}
\expandafter\ifx\csname url\endcsname\relax
  \def\url#1{\texttt{#1}}\fi
\expandafter\ifx\csname urlprefix\endcsname\relax\def\urlprefix{URL }\fi
\providecommand{\bibinfo}[2]{#2}
\providecommand{\eprint}[2][]{\url{#2}}

\bibitem{Kaufman2016}
\bibinfo{author}{Kaufman, A.~M.} \emph{et~al.}
\newblock \bibinfo{title}{{Quantum thermalization through entanglement in an
  isolated many-body system}}.
\newblock \emph{\bibinfo{journal}{Science}} \textbf{\bibinfo{volume}{353}},
  \bibinfo{pages}{794--800} (\bibinfo{year}{2016}).
\newblock \urlprefix\url{https://science.sciencemag.org/content/353/6301/794
  https://science.sciencemag.org/content/353/6301/794.abstract}.

\bibitem{Lukin2019}
\bibinfo{author}{Lukin, A.} \emph{et~al.}
\newblock \bibinfo{title}{{Probing entanglement in a many-body-localized
  system}}.
\newblock \emph{\bibinfo{journal}{Science}} \textbf{\bibinfo{volume}{364}},
  \bibinfo{pages}{256--260} (\bibinfo{year}{2019}).
\newblock
  \urlprefix\url{https://www-science-org.ezp-prod1.hul.harvard.edu/doi/abs/10.1126/science.aau0818}.
\newblock \eprint{1805.09819}.

\bibitem{Rispoli2019}
\bibinfo{author}{Rispoli, M.} \emph{et~al.}
\newblock \bibinfo{title}{{Quantum critical behaviour at the many-body
  localization transition}}.
\newblock \emph{\bibinfo{journal}{Nature 2019 573:7774}}
  \textbf{\bibinfo{volume}{573}}, \bibinfo{pages}{385--389}
  (\bibinfo{year}{2019}).
\newblock \urlprefix\url{https://www.nature.com/articles/s41586-019-1527-2}.

\bibitem{Morong2021}
\bibinfo{author}{Morong, W.} \emph{et~al.}
\newblock \bibinfo{title}{{Observation of Stark many-body localization without
  disorder}}.
\newblock \emph{\bibinfo{journal}{Nature |}} \textbf{\bibinfo{volume}{599}},
  \bibinfo{pages}{393} (\bibinfo{year}{2021}).
\newblock \urlprefix\url{https://doi.org/10.1038/s41586-021-03988-0}.

\bibitem{Zhang2017}
\bibinfo{author}{Zhang, J.} \emph{et~al.}
\newblock \bibinfo{title}{{Observation of a Discrete Time Crystal}}.
\newblock \emph{\bibinfo{journal}{Nature}} \textbf{\bibinfo{volume}{543}},
  \bibinfo{pages}{217} (\bibinfo{year}{2017}).
\newblock \eprint{1609.08684}.

\bibitem{Choi2017}
\bibinfo{author}{Choi, J.} \emph{et~al.}
\newblock \bibinfo{title}{state spin ensemble ( a ) ( b )}
  \textbf{\bibinfo{volume}{vi}}, \bibinfo{pages}{1--11} (\bibinfo{year}{2017}).

\bibitem{Keesling2019}
\bibinfo{author}{Keesling, A.} \emph{et~al.}
\newblock \bibinfo{title}{{Quantum Kibble–Zurek mechanism and critical
  dynamics on a programmable Rydberg simulator}}.
\newblock \emph{\bibinfo{journal}{Nature 2019 568:7751}}
  \textbf{\bibinfo{volume}{568}}, \bibinfo{pages}{207--211}
  (\bibinfo{year}{2019}).
\newblock \urlprefix\url{https://www.nature.com/articles/s41586-019-1070-1}.

\bibitem{Ebadi2021}
\bibinfo{author}{Ebadi, S.} \emph{et~al.}
\newblock \bibinfo{title}{{Quantum phases of matter on a 256-atom programmable
  quantum simulator}}.
\newblock \emph{\bibinfo{journal}{Nature 2021 595:7866}}
  \textbf{\bibinfo{volume}{595}}, \bibinfo{pages}{227--232}
  (\bibinfo{year}{2021}).
\newblock \urlprefix\url{https://www.nature.com/articles/s41586-021-03582-4}.

\bibitem{Kyprianidis2021}
\bibinfo{author}{Kyprianidis, A.} \emph{et~al.}
\newblock \bibinfo{title}{{Observation of a prethermal discrete time crystal}}.
\newblock \emph{\bibinfo{journal}{Science}} \textbf{\bibinfo{volume}{372}},
  \bibinfo{pages}{1192--1196} (\bibinfo{year}{2021}).
\newblock
  \urlprefix\url{https://www-science-org.ezp-prod1.hul.harvard.edu/doi/abs/10.1126/science.abg8102}.
\newblock \eprint{2102.01695}.

\bibitem{Turner}
\bibinfo{author}{Turner, C.~J.}, \bibinfo{author}{Michailidis, A.~A.},
  \bibinfo{author}{Abanin, D.~A.}, \bibinfo{author}{Serbyn, M.} \&
  \bibinfo{author}{Papi{\'{c}}, Z.}
\newblock \bibinfo{title}{{Weak ergodicity breaking from quantum many-body
  scars}} \urlprefix\url{https://doi.org/10.1038/s41567-018-0137-5}.

\bibitem{Bluvstein2021}
\bibinfo{author}{Bluvstein, D.} \emph{et~al.}
\newblock \bibinfo{title}{{Controlling quantum many-body dynamics in driven
  Rydberg atom arrays}}.
\newblock \emph{\bibinfo{journal}{Science}} \textbf{\bibinfo{volume}{371}},
  \bibinfo{pages}{1355--1359} (\bibinfo{year}{2021}).
\newblock
  \urlprefix\url{https://www-science-org.ezp-prod1.hul.harvard.edu/doi/abs/10.1126/science.abg2530}.

\bibitem{Kao2021}
\bibinfo{author}{Kao, W.}, \bibinfo{author}{Li, K.~Y.}, \bibinfo{author}{Lin,
  K.~Y.}, \bibinfo{author}{Gopalakrishnan, S.} \& \bibinfo{author}{Lev, B.~L.}
\newblock \bibinfo{title}{{Topological pumping of a 1D dipolar gas into
  strongly correlated prethermal states}}.
\newblock \emph{\bibinfo{journal}{Science}} \textbf{\bibinfo{volume}{371}},
  \bibinfo{pages}{296--300} (\bibinfo{year}{2021}).
\newblock
  \urlprefix\url{https://www-science-org.ezp-prod1.hul.harvard.edu/doi/abs/10.1126/science.abb4928}.

\bibitem{Cardellino2014}
\bibinfo{author}{Cardellino, J.} \emph{et~al.}
\newblock \bibinfo{title}{{The effect of spin transport on spin lifetime in
  nanoscale systems}}.
\newblock \emph{\bibinfo{journal}{Nature Nanotechnology}}
  \textbf{\bibinfo{volume}{9}}, \bibinfo{pages}{343--347}
  (\bibinfo{year}{2014}).

\bibitem{Alvarez2015}
\bibinfo{author}{Alvarez, G.~A.}, \bibinfo{author}{Suter, D.} \&
  \bibinfo{author}{Kaiser, R.}
\newblock \bibinfo{title}{{Localization-delocalization transition in the
  dynamics of dipolar-coupled nuclear spins}}.
\newblock \emph{\bibinfo{journal}{Science}} \textbf{\bibinfo{volume}{349}}
  (\bibinfo{year}{2015}).

\bibitem{Wei2018}
\bibinfo{author}{Wei, K.~X.}, \bibinfo{author}{Ramanathan, C.} \&
  \bibinfo{author}{Cappellaro, P.}
\newblock \bibinfo{title}{{Exploring Localization in Nuclear Spin Chains}}.
\newblock \emph{\bibinfo{journal}{Physical Review Letters}}
  \textbf{\bibinfo{volume}{120}}, \bibinfo{pages}{70501}
  (\bibinfo{year}{2018}).
\newblock
  \urlprefix\url{https://link.aps.org/doi/10.1103/PhysRevLett.120.070501}.

\bibitem{Smith2016}
\bibinfo{author}{Smith, J.} \emph{et~al.}
\newblock \bibinfo{title}{{Many-body localization in a quantum simulator with
  programmable random disorder}}.
\newblock \emph{\bibinfo{journal}{Nature Physics 2016 12:10}}
  \textbf{\bibinfo{volume}{12}}, \bibinfo{pages}{907--911}
  (\bibinfo{year}{2016}).
\newblock \urlprefix\url{https://www.nature.com/articles/nphys3783}.

\bibitem{Choi2016}
\bibinfo{author}{Choi, J.-y.} \emph{et~al.}
\newblock \bibinfo{title}{{Exploring the many-body localization transition in
  two dimensions.}}
\newblock \emph{\bibinfo{journal}{Science}} \textbf{\bibinfo{volume}{352}},
  \bibinfo{pages}{1547--1552} (\bibinfo{year}{2016}).
\newblock \urlprefix\url{http://www.ncbi.nlm.nih.gov/pubmed/27339981}.

\bibitem{Kucsko2018}
\bibinfo{author}{Kucsko, G.} \emph{et~al.}
\newblock \bibinfo{title}{{Critical Thermalization of a Disordered Dipolar Spin
  System in Diamond}}.
\newblock \emph{\bibinfo{journal}{Physical Review Letters}}
  \textbf{\bibinfo{volume}{121}}, \bibinfo{pages}{23601}
  (\bibinfo{year}{2018}).
\newblock
  \urlprefix\url{https://link.aps.org/doi/10.1103/PhysRevLett.121.023601}.

\bibitem{Schreiber2015}
\bibinfo{author}{Schreiber, M.} \emph{et~al.}
\newblock \bibinfo{title}{{Observation of many-body localization of interacting
  fermions in a quasi-random optical lattice}}  (\bibinfo{year}{2015}).
\newblock \urlprefix\url{http://arxiv.org/abs/1501.05661}.
\newblock \eprint{1501.05661}.

\bibitem{Feldman1996}
\bibinfo{author}{Fel'dman, E.~B.} \& \bibinfo{author}{Lacelle, S.}
\newblock \bibinfo{title}{{Cite as}}.
\newblock \emph{\bibinfo{journal}{J. Chem. Phys}}
  \textbf{\bibinfo{volume}{104}}, \bibinfo{pages}{2000} (\bibinfo{year}{1996}).
\newblock \urlprefix\url{https://doi.org/10.1063/1.470956}.

\bibitem{Dobrovitski2008}
\bibinfo{author}{Dobrovitski, V.~V.}, \bibinfo{author}{Feiguin, A.~E.},
  \bibinfo{author}{Awschalom, D.~D.} \& \bibinfo{author}{Hanson, R.}
\newblock \bibinfo{title}{{Decoherence dynamics of a single spin versus spin
  ensemble}}.
\newblock \emph{\bibinfo{journal}{Physical Review B - Condensed Matter and
  Materials Physics}} \textbf{\bibinfo{volume}{77}}, \bibinfo{pages}{245212}
  (\bibinfo{year}{2008}).
\newblock
  \urlprefix\url{https://journals-aps-org.ezp-prod1.hul.harvard.edu/prb/abstract/10.1103/PhysRevB.77.245212}.

\bibitem{Anderson1958}
\bibinfo{author}{Anderson, P.~W.}
\newblock \bibinfo{title}{{Absence of Diffusion in Certain Random Lattices}}.
\newblock \emph{\bibinfo{journal}{Physical Review}}
  \textbf{\bibinfo{volume}{109}}, \bibinfo{pages}{1492--1505}
  (\bibinfo{year}{1958}).

\bibitem{Burin2006}
\bibinfo{author}{Burin, A.~L.}
\newblock \bibinfo{title}{{Energy delocalization in strongly disordered systems
  induced by the long-range many-body interaction}}  (\bibinfo{year}{2006}).
\newblock \urlprefix\url{http://arxiv.org/abs/cond-mat/0611387}.
\newblock \eprint{0611387}.

\bibitem{Yao2014}
\bibinfo{author}{Yao, N.~Y.} \emph{et~al.}
\newblock \bibinfo{title}{{Many-body localization in dipolar systems}}.
\newblock \emph{\bibinfo{journal}{Physical Review Letters}}
  \textbf{\bibinfo{volume}{113}}, \bibinfo{pages}{243002}
  (\bibinfo{year}{2014}).
\newblock
  \urlprefix\url{https://journals.aps.org/prl/abstract/10.1103/PhysRevLett.113.243002}.

\bibitem{Gong2017a}
\bibinfo{author}{Gong, Z.-X.}, \bibinfo{author}{Foss-Feig, M.},
  \bibinfo{author}{Brand{\~{a}}o, F. G. S.~L.} \& \bibinfo{author}{Gorshkov,
  A.~V.}
\newblock \bibinfo{title}{{Entanglement Area Laws for Long-Range Interacting
  Systems}}.
\newblock \emph{\bibinfo{journal}{Physical Review Letters}}
  \textbf{\bibinfo{volume}{119}}, \bibinfo{pages}{50501}
  (\bibinfo{year}{2017}).
\newblock
  \urlprefix\url{http://link.aps.org/doi/10.1103/PhysRevLett.119.050501}.

\bibitem{Orioli2018}
\bibinfo{author}{Orioli, A.~P.} \emph{et~al.}
\newblock \bibinfo{title}{{Relaxation of an Isolated Dipolar-Interacting
  Rydberg Quantum Spin System}}.
\newblock \emph{\bibinfo{journal}{Physical Review Letters}}
  \textbf{\bibinfo{volume}{120}}, \bibinfo{pages}{063601}
  (\bibinfo{year}{2018}).
\newblock \eprint{1703.05957}.

\bibitem{Gopalakrishnan2016}
\bibinfo{author}{Gopalakrishnan, S.}, \bibinfo{author}{Agarwal, K.},
  \bibinfo{author}{Demler, E.~A.}, \bibinfo{author}{Huse, D.~A.} \&
  \bibinfo{author}{Knap, M.}
\newblock \bibinfo{title}{{Griffiths effects and slow dynamics in nearly
  many-body localized systems}}.
\newblock \emph{\bibinfo{journal}{Physical Review B}}
  \textbf{\bibinfo{volume}{93}}, \bibinfo{pages}{1--12} (\bibinfo{year}{2016}).
\newblock \eprint{1511.06389}.

\bibitem{Sushkov2014}
\bibinfo{author}{Sushkov, A.~O.} \emph{et~al.}
\newblock \bibinfo{title}{{Magnetic resonance detection of individual proton
  spins using quantum reporters}}.
\newblock \emph{\bibinfo{journal}{Physical Review Letters}}
  \textbf{\bibinfo{volume}{113}}, \bibinfo{pages}{197601}
  (\bibinfo{year}{2014}).
\newblock
  \urlprefix\url{https://journals-aps-org.ezp-prod1.hul.harvard.edu/prl/abstract/10.1103/PhysRevLett.113.197601}.

\bibitem{Davis2021}
\bibinfo{author}{Davis, E.~J.} \emph{et~al.}
\newblock \bibinfo{title}{{Probing many-body noise in a strongly interacting
  two-dimensional dipolar spin system}}  (\bibinfo{year}{2021}).
\newblock \urlprefix\url{http://arxiv.org/abs/2103.12742}.
\newblock \eprint{2103.12742}.

\bibitem{Grotz2011}
\bibinfo{author}{Grotz, B.} \emph{et~al.}
\newblock \bibinfo{title}{{Sensing external spins with nitrogen-vacancy
  diamond}}.
\newblock \emph{\bibinfo{journal}{New Journal of Physics}}
  \textbf{\bibinfo{volume}{13}}, \bibinfo{pages}{55004} (\bibinfo{year}{2011}).

\bibitem{Grinolds2014}
\bibinfo{author}{Grinolds, M.~S.} \emph{et~al.}
\newblock \bibinfo{title}{{Subnanometre resolution in three-dimensional
  magnetic resonance imaging of individual dark spins}}.
\newblock \emph{\bibinfo{journal}{Nature Nanotechnology}}
  \textbf{\bibinfo{volume}{9}}, \bibinfo{pages}{279--284}
  (\bibinfo{year}{2014}).
\newblock \urlprefix\url{www.nature.com/naturenanotechnology}.
\newblock \eprint{1401.2674}.

\bibitem{Tetienne2018}
\bibinfo{author}{Tetienne, J.~P.} \emph{et~al.}
\newblock \bibinfo{title}{{Spin properties of dense near-surface ensembles of
  nitrogen-vacancy centers in diamond}}.
\newblock \emph{\bibinfo{journal}{Physical Review B}}
  \textbf{\bibinfo{volume}{97}}, \bibinfo{pages}{085402}
  (\bibinfo{year}{2018}).
\newblock
  \urlprefix\url{https://journals-aps-org.ezp-prod1.hul.harvard.edu/prb/abstract/10.1103/PhysRevB.97.085402}.
\newblock \eprint{1711.04429}.

\bibitem{Sangtawesin2019}
\bibinfo{author}{Sangtawesin, S.} \emph{et~al.}
\newblock \bibinfo{title}{{Origins of Diamond Surface Noise Probed by
  Correlating Single-Spin Measurements with Surface Spectroscopy}}
  \textbf{\bibinfo{volume}{031052}}, \bibinfo{pages}{1--17}
  (\bibinfo{year}{2019}).

\bibitem{Stacey2019}
\bibinfo{author}{Stacey, A.} \emph{et~al.}
\newblock \bibinfo{title}{{Evidence for Primal sp <sup>2</sup> Defects at the
  Diamond Surface: Candidates for Electron Trapping and Noise Sources}}.
\newblock \emph{\bibinfo{journal}{Advanced Materials Interfaces}}
  \textbf{\bibinfo{volume}{6}}, \bibinfo{pages}{1801449}
  (\bibinfo{year}{2019}).
\newblock \urlprefix\url{http://doi.wiley.com/10.1002/admi.201801449}.

\bibitem{Rosskopf2014}
\bibinfo{author}{Rosskopf, T.} \emph{et~al.}
\newblock \bibinfo{title}{{Investigation of Surface Magnetic Noise by Shallow
  Spins in Diamond}} \textbf{\bibinfo{volume}{147602}}, \bibinfo{pages}{1--5}
  (\bibinfo{year}{2014}).

\bibitem{Romach2015}
\bibinfo{author}{Romach, Y.} \emph{et~al.}
\newblock \bibinfo{title}{{Spectroscopy of Surface-Induced Noise Using Shallow
  Spins in Diamond}}.
\newblock \emph{\bibinfo{journal}{Physical Review Letters}}
  \textbf{\bibinfo{volume}{114}}, \bibinfo{pages}{17601}
  (\bibinfo{year}{2015}).
\newblock
  \urlprefix\url{https://link.aps.org/doi/10.1103/PhysRevLett.114.017601}.

\bibitem{Myers2014}
\bibinfo{author}{Myers, B.~A.} \emph{et~al.}
\newblock \bibinfo{title}{{Probing Surface Noise with Depth-Calibrated Spins in
  Diamond}}.
\newblock \emph{\bibinfo{journal}{Physical Review Letters}}
  \textbf{\bibinfo{volume}{113}}, \bibinfo{pages}{27602}
  (\bibinfo{year}{2014}).

\bibitem{Myers2017}
\bibinfo{author}{Myers, B.~A.}, \bibinfo{author}{Ariyaratne, A.} \&
  \bibinfo{author}{Jayich, A.~B.}
\newblock \bibinfo{title}{{Double-Quantum Spin-Relaxation Limits to Coherence
  of Near-Surface Nitrogen-Vacancy Centers}}.
\newblock \emph{\bibinfo{journal}{Physical Review Letters}}
  \textbf{\bibinfo{volume}{118}}, \bibinfo{pages}{197201}
  (\bibinfo{year}{2017}).

\bibitem{Bluvstein2019}
\bibinfo{author}{Bluvstein, D.}, \bibinfo{author}{Zhang, Z.},
  \bibinfo{author}{McLellan, C.~A.}, \bibinfo{author}{Williams, N.~R.} \&
  \bibinfo{author}{Jayich, A.~C.}
\newblock \bibinfo{title}{{Extending the Quantum Coherence of a Near-Surface
  Qubit by Coherently Driving the Paramagnetic Surface Environment}}.
\newblock \emph{\bibinfo{journal}{Physical Review Letters}}
  \textbf{\bibinfo{volume}{123}}, \bibinfo{pages}{146804}
  (\bibinfo{year}{2019}).
\newblock
  \urlprefix\url{https://journals-aps-org.ezp-prod1.hul.harvard.edu/prl/abstract/10.1103/PhysRevLett.123.146804}.
\newblock \eprint{1905.06405}.

\bibitem{FavarodeOliveira2017}
\bibinfo{author}{{F{\'{a}}varo de Oliveira}, F.} \emph{et~al.}
\newblock \bibinfo{title}{{Tailoring spin defects in diamond by lattice
  charging}}.
\newblock \emph{\bibinfo{journal}{Nature Communications}}
  \textbf{\bibinfo{volume}{8}}, \bibinfo{pages}{15409} (\bibinfo{year}{2017}).
\newblock \urlprefix\url{http://www.nature.com/doifinder/10.1038/ncomms15409}.

\bibitem{Dwyer2021}
\bibinfo{author}{Dwyer, B.~L.} \emph{et~al.}
\newblock \bibinfo{title}{{Probing spin dynamics on diamond surfaces using a
  single quantum sensor}}  (\bibinfo{year}{2021}).
\newblock \urlprefix\url{http://arxiv.org/abs/2103.12757}.
\newblock \eprint{2103.12757}.

\bibitem{som}
 \emph{\bibinfo{title}{{Supplementary Information}}}.

\bibitem{Laraoui2013}
\bibinfo{author}{Laraoui, A.} \emph{et~al.}
\newblock \bibinfo{title}{{High-resolution correlation spectroscopy of 13C
  spins near a nitrogen-vacancy centre in diamond}}.
\newblock \emph{\bibinfo{journal}{Nature Communications}}
  \textbf{\bibinfo{volume}{4}}, \bibinfo{pages}{1--7} (\bibinfo{year}{2013}).
\newblock \urlprefix\url{www.nature.com/naturecommunications}.

\bibitem{Staudacher2015}
\bibinfo{author}{Staudacher, T.} \emph{et~al.}
\newblock \bibinfo{title}{{Probing molecular dynamics at the nanoscale via an
  individual paramagnetic centre}}.
\newblock \emph{\bibinfo{journal}{Nature Communications}}
  \textbf{\bibinfo{volume}{6}}, \bibinfo{pages}{8527} (\bibinfo{year}{2015}).

\bibitem{Loretz2014}
\bibinfo{author}{Loretz, M.} \emph{et~al.}
\newblock \bibinfo{title}{{Single-proton spin detection by diamond
  magnetometry}}.
\newblock \emph{\bibinfo{journal}{Science}}
  \bibinfo{pages}{science.1259464----} (\bibinfo{year}{2014}).

\bibitem{DeVience2015}
\bibinfo{author}{DeVience, S.~J.} \emph{et~al.}
\newblock \bibinfo{title}{{Nanoscale NMR spectroscopy and imaging of multiple
  nuclear species}}.
\newblock \emph{\bibinfo{journal}{Nature Nanotechnology}}
  \textbf{\bibinfo{volume}{10}}, \bibinfo{pages}{129--134}
  (\bibinfo{year}{2015}).
\newblock \urlprefix\url{http://www.nature.com/articles/nnano.2014.313}.

\bibitem{Abragam1961}
\bibinfo{author}{Abragam, A.}
\newblock \bibinfo{title}{{The Principles of Nuclear Magnetism: The
  International Series of Monographs on Physics}} (\bibinfo{year}{1961}).

\bibitem{Slichter1978}
\bibinfo{author}{Slichter, C.~P.}
\newblock \emph{\bibinfo{title}{{Principles of Magnetic Resonance}}},
  vol.~\bibinfo{volume}{1} of \emph{\bibinfo{series}{Springer Series in
  Solid-State Sciences}} (\bibinfo{publisher}{Springer Berlin Heidelberg},
  \bibinfo{address}{Berlin, Heidelberg}, \bibinfo{year}{1978}).
\newblock \urlprefix\url{http://link.springer.com/10.1007/978-3-662-12784-1}.

\bibitem{Graser2021}
\bibinfo{author}{Gr{\"{a}}{\ss}er, T.}, \bibinfo{author}{Bleicker, P.},
  \bibinfo{author}{Hering, D.-B.}, \bibinfo{author}{Yarmohammadi, M.} \&
  \bibinfo{author}{Uhrig, G.~S.}
\newblock \bibinfo{title}{{Dynamic mean-field theory for dense spin systems at
  infinite temperature}}  (\bibinfo{year}{2021}).
\newblock \urlprefix\url{https://arxiv.org/abs/2107.07821v1}.
\newblock \eprint{2107.07821}.

\bibitem{Taylor2008a}
\bibinfo{author}{Taylor, J.~M.} \emph{et~al.}
\newblock \bibinfo{title}{{High-sensitivity diamond magnetometer with nanoscale
  resolution}}.
\newblock \emph{\bibinfo{journal}{Nature Physics}}
  \textbf{\bibinfo{volume}{4}}, \bibinfo{pages}{810--816}
  (\bibinfo{year}{2008}).
\newblock \urlprefix\url{www.nature.com/naturephysics}.
\newblock \eprint{0805.1367}.

\bibitem{Mamin2013}
\bibinfo{author}{Mamin, H.~J.} \emph{et~al.}
\newblock \bibinfo{title}{{Nanoscale nuclear magnetic resonance with a
  nitrogen-vacancy spin sensor.}}
\newblock \emph{\bibinfo{journal}{Science}} \textbf{\bibinfo{volume}{339}},
  \bibinfo{pages}{557--560} (\bibinfo{year}{2013}).

\bibitem{Pham2016}
\bibinfo{author}{Pham, L.~M.} \emph{et~al.}
\newblock \bibinfo{title}{{NMR technique for determining the depth of shallow
  nitrogen-vacancy centers in diamond}}.
\newblock \emph{\bibinfo{journal}{Physical Review B}}
  \textbf{\bibinfo{volume}{93}} (\bibinfo{year}{2016}).
\newblock \eprint{1508.04191}.

\end{thebibliography}
 
\end{document}